\documentclass[12pt]{article}

\def\bfu{\mbox{\bf u}}
\def\bfB{\mbox{\bf B}}
\def\bfBstar{\mbox{\bf B$^\star$}}
\def\bfustar{\mbox{\bf u$^\star$}}
\def\Tstar{T^\star}
\def\bfe{\mbox{\bf e}}
\def\bfr{\mbox{\bf r}}
\def\dd{{\rm d}}
\def\bfnabla{\mbox{\boldmath $\nabla$}}

\newcommand*{\Ray}{{\rm Ra}}
\newcommand*{\Power}{{\rm P}}
\newcommand*{\Pra}{{\rm Pr}}
\newcommand*{\Pm}{{\rm Pm}}
\newcommand*{\Rey}{{\rm Re}}
\newcommand*{\Rm}{{\rm Rm}}
\newcommand*{\q}{{\rm q}}
\newcommand*{\Ek}{{\rm E}}
\newcommand*{\Nu}{{\rm Nu}}
\newcommand*{\Lo}{{\rm Lo}}
\newcommand{\Ro}{{\rm Ro}}

\newcommand{\chirel}{{\chi_{\rm rel}}}
\newcommand{\tildel}{{\tilde{\ell}}}
\newcommand{\B}{{\rm B}}
\newcommand{\uu}{{\rm u}}

\usepackage{vmargin}                                
\usepackage[utf8]{inputenc}
\usepackage[T1]{fontenc}
\usepackage{color}
\usepackage{graphicx}
\usepackage{makeidx}                          
\usepackage{amsmath,amssymb}
\usepackage{natbib} 
\newcommand{\X}{\! \times \!}
\newcommand{\fohm}{{f_{\rm ohm}}}
\newcommand{\fdip}{{f_{\rm dip}}}

\setmarginsrb{1.5cm}{0.5cm}{1.5cm}{2.5cm}{0cm}{1cm}{0cm}{1cm}   

\begin{document}

\title{Predictive Scaling Laws for Spherical Rotating Dynamos} 

\author{L. Oruba and E. Dormy}

\date{}

\maketitle

\begin{abstract}

State of the art numerical models of the Geodynamo are still performed in a parameter regime extremely 
remote from the values relevant to the physics of the Earth's core. In order to establish a connection 
between dynamo modeling and the geophysical motivation, {it is necessary to use} 
scaling laws. Such scaling laws 
establish the dependence of essential quantities (such as the magnetic field strength) on measured or 
controlled quantities. 
They allow for a direct confrontation of advanced models with geophysical {constraints}.

We combine a numerical approach, based on a multiple linear regression
method in the form of power laws, applied to  a database of $102$ direct
numerical simulations (courtesy of U. Christensen), 
and a physical approach, based on energetics and forces balances.

We show that previous empirical scaling laws for the magnetic field
strength essentially reflect the statistical balance between energy
production and dissipation for saturated dynamos.
Such power based scaling laws are thus necessarily valid for any dynamo in statistical equilibrium
and applicable to any numerical model, irrespectively of the dynamo mechanism.

We show that direct numerical fits can provide contradictory results
owing to biases in the parameters space covered in the numerics and to the
role of a priori hypothesis on the fraction of 
ohmic dissipation.

We introduce predictive scaling laws, i.e. relations involving input
parameters of the governing equations only. We guide our reasoning on
physical considerations. 
We show that our predictive scaling laws can properly describe the numerical database and
reflect the dominant forces balance at work in these numerical simulations. We
highlight the dependence of the magnetic field strength on the rotation rate. 
Finally, our results stress that available numerical models
operate in a viscous dynamical regime, which is not relevant to the Earth's core.

\end{abstract} 

\section{Introduction}

\label{Intro}
Many numerical models have been produced over the last few years to try and
reproduce characteristics of planetary and stellar magnetic fields.
The parameter regime relevant to these natural objects is however out of
reach of present days computational resources. In order to assess the
reliability of current numerical models and their relevance to natural
applications, it is thus necessary to rely on scaling laws, which can be
established on the basis of a set of numerical models with varying
control parameters and then extended to the regime of geophysical or
astrophysical relevance. 

Previous empirical scaling laws for the magnetic field strength
\citep{Chr06} have proven
to be remarkably robust. Indeed they seem to be applicable to numerical
models irrespectively of the parameter regime, viscous or inertial \citep{Chr10,SPD12},
as well as to natural objects of very different kinds \citep{Chr09}.
Such scaling laws are constructed on the basis of a statistical balance between energy
production and dissipation. It is essential to separate the relative
importance of this general assumption --which will necessarily be valid for
any dynamo in statistical equilibrium-- from additional assumptions which
could test the nature of a particular dynamo. An additional key issue is that such
existing relations only relate measured quantity. They have no predictive
power for numerical models in the sense that the knowledge of control
parameters (entering the governing equations) is not sufficient to a priori
estimate the strength of the produced magnetic field. We therefore want to
introduce predictive scaling laws, which a priori estimate the amplitude of
a measured quantity (say the magnetic field strength) as a function of
input parameters only.

\section{Governing equations and numerical models}
\label{equations}
We restrict our study to Boussinesq models of planetary dynamos. The domain consists of a spherical shell, and the aspect 
ratio between the two bounding spheres is set to $\xi\equiv r_i/r_o=0.35 \,$. The flow is driven by an
imposed difference of temperature between the inner and outer boundaries.

The governing equations in the rotating reference frame can then be
written -- using $L=r_o-r_i$ as unit of length, $\Omega ^{-1}$ as unit of
time, $\Delta T$ as unit of temperature, and $\sqrt{\rho \mu} \, \Omega L$ as
unit for the magnetic field -- as
\begin{equation}
\partial _t \bfustar +  (\bfustar \cdot \bfnabla) \bfustar 
=
- \bfnabla \pi^{\star} 
+ {\rm E} \, \Delta \bfustar
- 2 \bfe_z \times \bfustar 
+ \frac{{\rm Ra}\, {\rm E}^2}{\rm Pr} \, \Tstar \, \frac{\bfr}{r_o}
+ \left(\bfnabla \times \bfBstar \right) \times \bfBstar\, , \,\,\,\,\,\,
\label{eq_NS}
\end{equation} 
\begin{equation}
\partial _t \bfBstar = \bfnabla \times (\bfustar \times \bfBstar) 
+ \frac{\rm E}{\rm Pm} \, \Delta \bfBstar \, ,
\qquad
\partial_t \Tstar + (\bfustar \cdot \bfnabla) \Tstar
= \frac{\rm E}{\rm Pr} \Delta \Tstar\, ,
\label{eq_ind}
\end{equation} 
\ \vskip -6mm
\begin{equation}
\bfnabla \cdot \bfustar = 
\bfnabla \cdot \bfBstar = 
0\, .
\label{eq_div}
\end{equation} 
Because the governing equations involve nine independent physical
parameters ($\alpha,\, g_0,\, \Delta T , \, \nu , \, \kappa , \, \eta, \,
\Omega ,$ $\rho, \, \mu  $) and five units (kg, m, s, K, C), owing to the Buckingham $\pi$ theorem, only
four independent non-dimensional parameters can be introduced.
In our system~(\ref{eq_NS}--\ref{eq_div}), they are 
the Ekman number $\Ek = \nu / (\Omega L ^2) \, ,$
the Prandtl number ${\rm Pr}={\nu}/{\kappa}\, , $ 
the magnetic Prandtl number ${\rm Pm}={\nu}/{\eta}\, ,$
and the Rayleigh number $\Ray = \alpha g_0 \Delta T L ^3 / (\nu \kappa)\, ,$
in which $\nu$ is the kinematic viscosity of the fluid, 
$\alpha$ the coefficient of thermal expansion, $g_0$ the gravity at the outer bounding sphere, 
$\kappa=k/(\rho c)$ its thermal diffusivity, 
and $\eta$ its magnetic diffusivity.
Throughout this article, non-dimensional quantities are denoted with a $^\star$.

All the simulations used in this work rely on no-slip mechanical boundary
conditions and an insulating outer domain. The inner core is
insulating in most simulations, and a few simulations involve a conducting inner core 
with the same conductivity as the fluid.

Our analysis will be tested against a 
wide database of 185 direct numerical simulations kindly provided by
U.~Christensen. 
The data sample 
is reduced to $102$ to only take into account dynamo simulations corresponding to fully developed 
convection ($\Nu>2$) and producing a dipolar magnetic field (relative dipole field strength 
$\fdip$ larger than $0.5$). Moreover, we limit our study to 
$\Pra \leq 10$, that is to say to values not too far from the value estimated for the Earth's core.
We will also highlight the subset of this database which was used 
in \cite{Chr06}. It is composed of $65$ runs available at the time.  
Finally we will use a few additional 
numerical data published in \cite{Morin09}.

These numerical data can be used to test scaling laws guided by physical
reasoning. It can also be used to establish direct numerical fits.
To this end, we introduce a multiple linear regression approach 
\citep{Mont01,Cornillon10}, detailed in appendix~\ref{mlr}.

\section{Power based scalings, key parameters and their relations}

\subsection{Energy balance between production and dissipation}
\label{ProdDiss}

In order to derive a scaling law for the magnetic field strength,
a possible approach introduced by \cite{Chr06} is to consider the statistical balance between energy
production by buoyancy forces and dissipation.
Time averaged quantities of a statistically steady dynamo state
should obviously satisfy
\begin{equation}
\Power =  D_{\eta}+D_{\nu} \, ,
\end{equation} 
where $\Power$ is the power generated by buoyancy forces,
$D_{\eta}$ is the rate of ohmic dissipation
\[
D_{\eta}= \int_{V}
\frac{\eta}{\mu} \, (\mathbf{\nabla} \X \mathbf{B})^2\, \dd V \, ,
\qquad \mbox{i.e.} \quad
D_{\eta}^{\star}= \Ek_{\eta} \int_{V} (\mathbf{\nabla} \X
\mathbf{B^\star})^2 \,\dd V^\star \, ,
\] 
in which $\Ek_{\eta}=\Ek/\Pm$ is the magnetic Ekman number and $D_{\nu}$ is the rate of viscous
dissipation 
\[
D_{\nu}= \int_{V} \rho \nu \, (\mathbf{\nabla} \X
\mathbf{u})^2\, \dd V \, ,
\qquad \mbox{i.e.} \quad
D_{\nu}^{\star}= \Ek \int_{V} (\mathbf{\nabla} \X
\mathbf{u^\star})^2\, \dd V^\star 
\, .
\]
The above quantities are all defined as
time averaged over a sufficient amount of time, so that they are steady for
a given parameter set.

Following \cite{Chr06} and introducing the $\fohm$ coefficient, defined as
\begin{equation} 
\fohm \equiv\frac{D_{\eta}}{D_{\eta}+D_{\nu}} \, ,
\label{def_f_ohm}
\end{equation} 
we get
\begin{equation} 
\Power = \frac{D_{\eta}}{\fohm} 
= \frac{1}{\fohm}\, \int_{V} \frac{\eta}{\mu} \, (\mathbf{\nabla} \times \mathbf{B})^2 \,\dd V
\sim \frac{1}{\fohm}\, \frac{\eta}{\mu} \,\frac{\B^2}{\ell_B^2}\, V \, ,
\end{equation} 
where we introduced a typical magnetic field strength $\B$ and a magnetic dissipation length scale $\ell_B$,
defined again using time averaged quantities as
\begin{equation} 
\ell_B ^2\equiv \frac{\int_V \mathbf{B}^2 \,\dd V}{\int_V (\mathbf{\nabla} \times \mathbf{B})^2\, \dd V}
= 2 \, \eta \, \frac{\Ek_{\rm mag}}{D_\eta} \qquad \mbox{i.e.} \qquad {\ell_B^\star}^2 \equiv 2\, \Ek_{\eta}\, \frac{\Ek_{mag}^{\star}}{D_{\eta}^{\star}} \, ,
\label{deflB}
\end{equation} 
\begin{equation} 
\mbox{with}\qquad
\Ek_{mag}=
\int_{V} \frac{\mathbf{B}^2}{2 \mu} \,\dd V \, , \qquad \mbox{i.e.}\qquad \Ek_{mag}^{\star}= \int_{V}
\frac{{\mathbf{B}^{\star}}^2}{2} \,\dd V^\star \, .
\end{equation} 
This simple reasoning provides the following expression for the magnetic
field strength 
\begin{equation} 
\frac{\B^2}{\mu} \sim \fohm \, \ell_B^2 \, \frac{\Power}{\eta \,V} 
= \fohm \, \ell_B^2 \, \frac{\rho \, \Power_M}{\eta} \, ,
\label{eqB}
\end{equation} 
where $\Power_M$ is the mass power generated by buoyancy forces
$\Power_M\equiv \Power/(\rho V)$~.

The non-dimensional form of equation (\ref{eqB}) is
\begin{equation}
\Lo \sim \fohm^{1/2}\, {\Power^\star}^{1/2} \,\Ek_{\eta}^{-1/2}\, {\ell_B^\star} \, ,
\label{eqLoadim}
\end{equation}
where $\Lo\equiv\left(2\, {\Ek_{mag}}^\star/V^\star\right)^{\frac{1}{2}}
\equiv \B^\star$. 
Expressing a scaling law for $\B$ (or its non-dimensional form $\Lo$) therefore reduces
to relating $\Power$ and $\ell_B$  to the relevant parameters.

In previous studies $\ell_B$ has often not been
introduced as such \citep[but see the review by][]{RobertsKing13}. 
Instead it is usually indirectly evaluated by introducing
the magnetic dissipation time $\tau_{\rm diss}\equiv \Ek_{mag}/D_{\eta}
=\ell_B^2/(2 \eta)\,$ \citep[see][]{CT04}, or in non-dimensional form
\(
\tau_{\eta}^\star
\equiv \tau_{diss}/\tau_{dip}\, ,
\) 
where
$\tau_{\rm dip} \equiv L^2/(\pi^2 \, \eta) \,$. This definition leads to
\(\tau_{\eta}^\star = \pi^2/2 \,\, {\ell_B^{\star}}^2\)~.
Besides, the parameter $\fohm$ is a rather complex number, which involves both a priori input and a posteriori output model
properties. It is usually assumed to be order one in natural dynamos \citep[but see][]{Schrinner12}, 
its importance in scaling laws is discussed in appendix~\ref{Appfohm}.

\subsection{Power generated by buoyancy forces}
\label{Power_section}
\cite{Chr06} established a relation between $\Power^\star$ and a
flux-based Rayleigh number $\Ray_Q^\star$
\begin{equation}
\Ray_Q^\star \equiv \frac{1}{4 \pi \,r_o\, r_i} \, \frac{\alpha \,g\, r_o \,\Delta Q}{\rho
  \, c \, \Omega^{3} \, (r_o-r_i)^2}\, ,
\label{def0_RastarQ}
\end{equation}
where $\Delta Q$ is the difference between the time-average total heat flow $Q$
and $Q_d^{T_s} = 4 \pi k T_a$ ($J\cdot s^{-1}$), which corresponds to the
diffusive heat flow associated to $T_S(r)= {T_a}/{r}+T_b$.\\
They show that 
\begin{equation}
\Power^\star \approx 2 \pi\, \xi\, \frac{1+\xi}{\left(1-\xi\right)^2} \,\Ray_Q^\star\, ,
\label{Power_CA06}
\end{equation}
under the assumptions that 
the volume integral of the realised temperature gradient can be
approximated by its conductive counterpart.
The demonstration requires in particular fixed
temperature boundary conditions. Relation (\ref{Power_CA06}) 
is well-verified for the numerical database used in the present study. That is why in the following 
numerical analysis of scaling laws, the parameter $\Power^\star$ will be replaced by $\Ray_Q^\star$ with a
prefactor of $7.03$ determined by the geometry via the aspect ratio $\xi$.

It is important to stress that \(\Ray_Q^\star\) is an output parameter, and
cannot be controlled a priori when using fixed temperature boundary
conditions.
It can however be related to the classical Rayleigh number,
 which is a control
parameter of the problem.
Indeed, introducing the Nusselt number $\Nu \equiv {Q}/{Q_d^{T_s}}$, which can be rewritten as $\Nu={Q_d(r_o)}/(4 \pi k T_a)$ 
under the statistically steady assumption, 
relation (\ref{def0_RastarQ}) becomes   
\begin{equation}
\Ray_Q^\star= \Ek^3\,\Pra^{-2} \, \Ray \, (\Nu-1) \, .
\label{def_RastarQ}
\end{equation}

The $\Ray_Q^\star$ parameter can not be controlled in the problem because
it is related to the output parameter $\Nu$. Its value is zero at the
onset of convection ($\Nu=1$). 

Note that \( \Ray_Q^\star \, \Nu / (\Nu-1) \) would be an input control
parameter in the case of imposed heat flux boundary conditions. The
construction of \(\Ray_Q^\star\) would however still involve, even with
such boundary conditions, measurements of the Nusselt number, because the
temperature difference accross the shell becomes a measured quantity.

\subsection{Role of the magnetic dissipation length scale $\ell_B$}
\label{lengthb}
\begin{figure}
\centerline{\includegraphics[width=0.42 \textwidth]{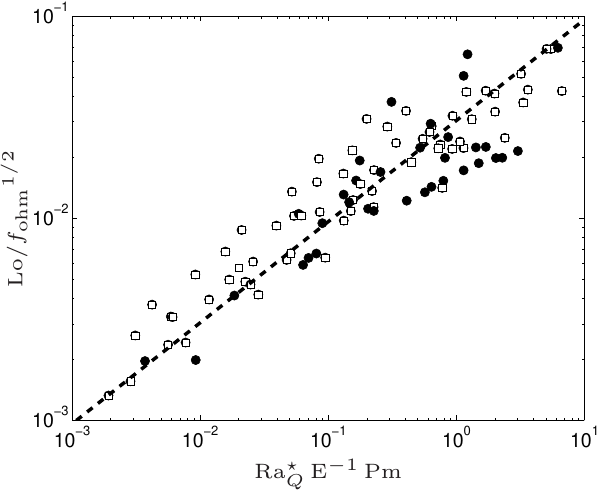}}
\caption{The Lorentz number corrected for the relative fraction of Ohmic dissipation versus a combination of 
the flux-based Rayleigh number, the Ekman number and the magnetic Prandtl number: equation (\ref{eqLoadim_lBcst}). 
This simple scaling law only reflects the statistical balance between energy production and dissipation, combined 
with a constant $\ell_B$. Points correspond to 
the full $102$ dynamos database, open squares indicate the subset used in \cite{Chr06}.}
\label{figsimple}
\end{figure}

In the numerical database used in the present paper, the dissipation length scale $\ell_B^{\star}$, calculated
using equation (\ref{deflB}), varies between $0.02$ and $0.10$. These values are obviously
smaller than those corresponding to the pure dipole decay in the absence of motions $\tau_{\eta}^\star=1/2$,
i.e. $\ell_B^{\star}=1/\pi \simeq 0.30$. Besides, the range of variation of $\ell_B^{\star}$ is less than one order of magnitude.
Thus, as a first approximation, the variations of $\ell_B^{\star}$ can be neglected, namely it can be set to a constant 
in equation (\ref{eqLoadim}). Using the relation (\ref{Power_CA06}), equation (\ref{eqLoadim}) becomes under this assumption 
\begin{equation}
\Lo \sim \fohm^{1/2} {\Ray_Q^\star}^{1/2} \,\Ek_{\eta}^{-1/2} \, .
\label{eqLoadim_lBcst}
\end{equation}
Its application to the $102$ dynamos database is represented in
figure~\ref{figsimple}, and yields the relative misfit $\chirel=0.433$.
Relation (\ref{eqLoadim_lBcst}), which simply corresponds to the energy balance between 
production and dissipation with $\ell_B$ approximated as a constant,
already provides a good fit to the numerical data. 
This implies that empirical fits of the magnetic field strength previously
obtained in the literature mainly reflect this simple balance
between energy production and dissipation, combined with an improved 
description of the magnetic dissipation $\ell_B$ than a simple constant,
which is however not essential.

The statistical balance between both terms of the right-hand side of the dimensional
form of the induction equation (\ref{eq_ind}) 
yields to $\uu\B/\ell \sim \eta \B /\ell_B^2\,$, where we introduced a typical velocity field strength $\uu$,
and $\ell$ has the dimension of a length scale which
depends on correlations between the norm and direction of $\bfu$ and
$\bfB$. The length scales $\ell_B$ and $\ell$ are thus related by  
\begin{equation}
\ell_B \sim \eta^{1/2} \, \uu^{-1/2}\, \ell^{1/2} \, ,
\label{eqlBlu}
\end{equation}
which can be normalised as
\begin{equation}
{\ell_B^\star} \sim \Rm^{-1/2} \,{\ell^\star}^{1/2} \, , \qquad \mbox{or}
\qquad {\ell_B^\star} \sim \Ek_{\eta}^{1/2} \,\Ro^{-1/2}\,{\ell^\star}^{1/2}\, ,
\label{eql}
\end{equation}
where $\Rm$ is the magnetic Reynolds number, and $\Ro$ is the Rossby
number, defined as
\(
\Ro \equiv \left(2\,{\Ek_{kin}}^\star/V^\star\right)^{\frac{1}{2}} \equiv
u^\star\, ,
\) 
\begin{equation} 
\mbox{with}\qquad \Ek_{\rm kin}\equiv \int_{V} \frac{\rho\, \bfu^2}{2}\, \dd V
\qquad \mbox{i.e.}\qquad
\Ek_{\rm kin}^{\star}\equiv \int_{V} \frac{{\bfu^{\star}}^2}{2}\, \dd V^\star \, .
\end{equation} 
The magnetic dissipation length scale $\ell_B$ is thus an output parameter, in
so far as it is related to both the characteristic velocity $\uu$ of the flow
(measured by $\Ro$ or $\Rm$) and the length scale $\ell$ (see appendix~\ref{section_lB_Rm}).

\subsection{Existing scaling laws for the magnetic field strength and their physical interpretation}
\label{inter_phys}

\begin{figure}
\centerline{\includegraphics[width=0.9 \textwidth]{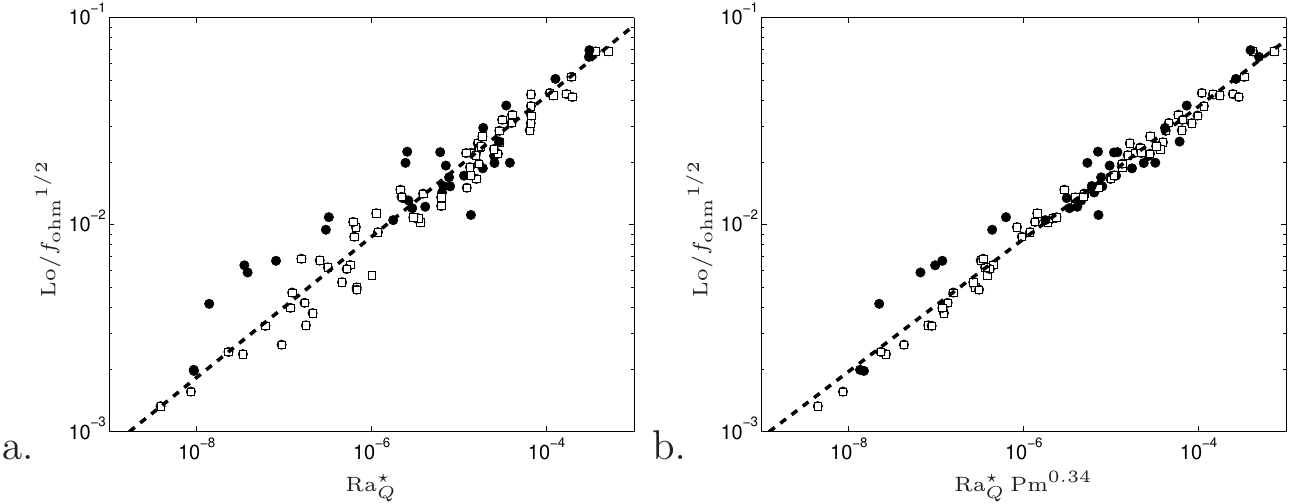}}
\caption{The Lorentz number corrected for the relative fraction of Ohmic dissipation versus a combination of 
the flux-based Rayleigh number and the magnetic Prandtl number, as proposed by 
 \cite{Chr06}: (a) relation (\ref{fitCA06_nonopt}), (b) relation
 (\ref{fitCA06}).
 Blacks points correspond to the $102$ dynamos database, open squares
 indicate the subset of data used in \cite{Chr06}.}
\label{FigloiUCR}
\end{figure}

\cite{Chr06} introduced two seminal scaling laws
\begin{equation}
{\Lo} \sim {\fohm^{1/2}}\,{\Ray_Q^\star}^{0.34}\,,
\label{fitCA06_nonopt}
\end{equation} 
and its optimised form
\begin{equation}
{\Lo} \sim  {\fohm^{1/2}}\,{\Ray_Q^\star}^{0.32} \,\Pm^{0.11}\, .
\label{fitCA06}
\end{equation} 
Their application to the $102$ dynamos database is represented in figure~\ref{FigloiUCR}, 
and yields the relative misfits
$\chirel=0.256 \, ,$ and $\ \chirel=0.152 \,$ respectively.
The corresponding assumption on the magnetic dissipation length scale ${\ell_B^\star}$ is detailed in {appendix~\ref{range_ra_star_Q}}. 
It respectively yields
\begin{equation}
{\ell_B^\star} \sim {\Ray_Q^\star}^{-0.16}\, \Ek_{\eta}^{1/2} \,, \qquad \mbox{and}
\qquad
{\ell_B^\star} \sim {\Ray_Q^\star}^{-0.18}\, \Ek^{1/2} \, \Pm^{-0.39}\, ,
\label{lB_CA06_eq}
\end{equation}
which are represented in figure~\ref{FigCA06} (see
appendix~\ref{range_ra_star_Q} for discussion).

\begin{figure}
\centerline{\includegraphics[width=0.9 \textwidth]{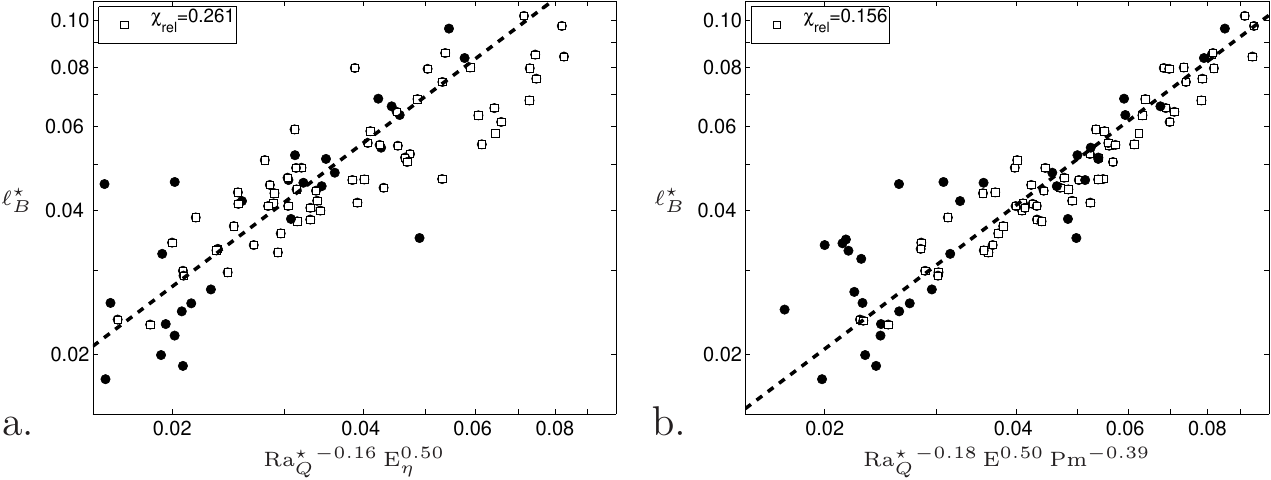}}
\caption{The magnetic dissipation length scale versus a combination of the flux-based Rayleigh number, the Ekman number
and the magnetic Prandtl number, as implied by \cite{Chr06} results (relations~(\ref{lB_CA06_eq})).
Black points correspond to the full $102$ dynamos database, open squares indicate 
the subset used in \cite{Chr06}. 
}
\label{FigCA06}
\end{figure}

Relation (\ref{fitCA06_nonopt}) and its optimised form (\ref{fitCA06}) are
empirical laws  
obtained using numerical experiments. The physical interpretation of relation (\ref{fitCA06_nonopt}), as
provided by \cite{Chr06}, is based on two assumptions: 
the empirical scaling law for the magnetic dissipation time $\tau_{\eta}^\star \sim \Rm^{-1}$ 
\citep{CT04}, which is equivalent to assuming $\ell^\star \sim 1$ {(see appendix~\ref{section_lB_Rm})},
and their empirical fit $\Ro \sim {\Ray_Q^\star}^{0.41}$
\citep[equation~(30) in][]{Chr06}. Using equation (\ref{eql}),
these two assumptions provide ${\ell_B^\star} \sim
{\Ray_Q^\star}^{-0.21}\,\Ek_{\eta}^{1/2}$. This last expression can then be injected in equation (\ref{eqLoadim}),
to yield ${\Lo} \sim \fohm^{1/2} \, {{\Ray_Q^\star}}^{0.29}$. Thus, their demonstration leads to an exponent of $\Ray_Q^\star$ 
equal to $0.29$, which is only slightly lower than their optimal exponent $0.34$ in (\ref{fitCA06_nonopt}).

\cite{Chr10} proposed a modified interpretation: while retaining the
assumption $\ell^\star \sim 1$, he replaced the scaling law for $\Ro$ by the 
one resulting from mixing length theory (balance between inertia and buoyancy). This theory, usually applied for turbulent 
convection in stars \citep{Stevenson79,Kippen90}, provides $\Ro \sim {\Ray_Q^\star}^{1/3}$. It leads to ${\Lo} \sim 
{\fohm}^{1/2} \, {\Ray_Q^\star}^{1/3}$, 
which is closer to the original fit (\ref{fitCA06_nonopt}) obtained by
\cite{Chr06}. 
Instead, \cite{Jones11} based his 
physical reasoning on the inertial scaling law $\Ro \sim {\Ray_Q^\star}^{2/5}$ \citep[derived 
from the so-called IAC balance, see][]{Aubert01}, 
and obtained ${\Lo} \sim {\fohm}^{1/2} \, {\Ray_Q^\star}^{0.30}$. The
assumptions of inertial scaling laws for $\Ro$ made by both \cite{Chr10}
and \cite{Jones11} however do not seem relevant to 
dipolar 
numerical dynamos \citep[see section~\ref{Ro_section} of this paper; and][]{Chr06,Soderlund12}. 

More recently, \cite{Davidson13} studied analytically the asymptotic limit expected to be relevant 
to planetary dynamos. In this limit, viscosity is negligible, which implies a vanishing viscous dissipation
($\fohm\sim 1$), and inertial forces do not enter the dominant forces balance 
(small Rossby number limit). Davidson's argument relies on a
dimensional analysis. 
On the right-hand side of equation (\ref{eqB}), with $\fohm= 1$,  both
$\Power_M$ and ${\ell_B^2}/\eta$ are assumed to be independent on
$\Omega$. This implies that ${B^2}/(\rho \mu)$ only depends on $L$ and
$\Power_M$, and thus 
\begin{equation}
\frac{\B^2}{\rho \,\mu} \sim L^{2/3}\,{\Power_M}^{2/3} \, ,
\label{Davidson1}
\end{equation} 
\citep[see equation~(6) in][]{Davidson13}. 
In order to account for viscous effects in numerical simulations, \cite{Davidson13}
 then replaces $\Power_M$ with $\fohm \Power_M$ in (\ref{Davidson1}),
which leads to  
\begin{equation}
\frac{\B^2}{\rho\, \mu} \sim L^{2/3}\,({\fohm \,\Power_M})^{2/3} \, ,
\label{eq9Dav}
\end{equation}
\citep[equation~(9) in][]{Davidson13}. It can be rewritten in its non-dimensional form as
\begin{equation}
{\Lo} \sim {\fohm^{1/3}}\,{\Ray_Q^\star}^{1/3} \, .
\label{eq9Dav_adim}
\end{equation}

\begin{figure}
\centerline{\includegraphics[width=0.42 \textwidth]{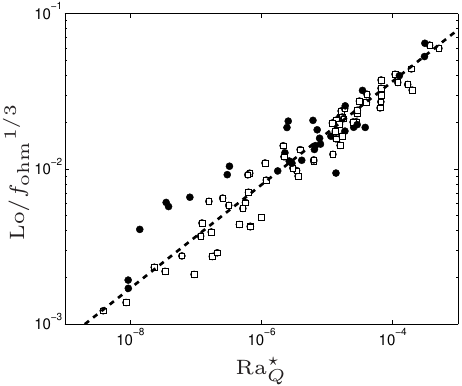}}
\caption{The Lorentz number corrected for the relative fraction of Ohmic dissipation versus a combination of 
the flux-based Rayleigh number and the magnetic Prandtl number, as proposed by 
\cite{Davidson13}, relation (\ref{eq9Dav_adim}). Blacks points correspond
to the $102$ dynamos database, open squares 
 indicate the subset of data used in \cite{Chr06}.}
\label{FigloiDavidson}
\end{figure}

Note that relation (\ref{eq9Dav_adim}) is based on physical considerations 
valid for the Earth's core but not necessarily realised in direct numerical
simulations (see appendix~\ref{AppDav}). 
It is similar to (\ref{fitCA06_nonopt}) except for the exponent of $\fohm$. 
The importance of this measured quantity in the efficiency of the power based
scaling laws is investigated in appendix~\ref{Appfohm}.
Its application to the $102$ dynamos database is represented in figure~\ref{FigloiDavidson}, 
and yields a relative misfit $\chirel=0.286$.

We discussed above three scaling laws proposed for the magnetic field
strength primarily as a function of the available power generated by
buoyancy forces and corresponding to equations (\ref{fitCA06_nonopt}), (\ref{fitCA06}) and 
(\ref{eq9Dav_adim}). 
Their application to our dynamos database is represented in
figures~\ref{FigloiUCR} and \ref{FigloiDavidson}.
Note that extending the $65$ dynamos database of \cite{Chr06} to the $102$ dynamos
database provided by U.~Christensen and used in the present paper, leads to a lower quality fit for the magnetic field amplitude 
\citep[compare figures~8-9 in][with figures~\ref{FigloiUCR}.a,b in the present paper]{Chr06}.
The three relations offer a good description of the available
numerical data, with relative misfits between $0.15$ and $0.30$. The best one is naturally relation (\ref{fitCA06}), 
since it involves a supplementary parameter $\Pm$ compared to scaling laws (\ref{fitCA06_nonopt}) 
and (\ref{eq9Dav_adim}). 

It is interesting to compare these three relations with the most simple form which stems from the energy
balance between production and dissipation and the assumption that $\ell_B$ is constant (dominant dipole field). This expression 
is represented in figure~\ref{figsimple} (see also equation (\ref{eqLoadim_lBcst})).
The relative misfit is only improved by some $50\%$ from this last relation to relations (\ref{fitCA06_nonopt}), 
(\ref{fitCA06}) and (\ref{eq9Dav_adim}) which all attempt to a finer description of the magnetic dissipation
length scale. The range of variation of $\ell_B$ in numerical models is necessarily restricted between 
the discretisation size and the size $L$ of the model. The key assumption is thus the statistical balance 
between energy production and dissipation, which is bound to work 
for any statistically steady dynamo (as illustrated in figure~\ref{figsimple}). This explains why the power based scaling 
law (\ref{fitCA06_nonopt}) was found to work 
with different prefactors for dipolar and multipolar dynamos, despite of their different induction mechanisms \citep{Chr10,SPD12}. 

\section{Predictive scaling laws for the magnetic field strength}

Power based scaling laws, discussed in the previous section, properly
describe the numerical database. However they only relate together measured
quantities.
We now want to express scaling laws
which only involve input parameters on the right-hand side. Such scaling laws 
will be referred to as ``predictive'' in the sense that 
they estimate the strength of a measured quantity, say the magnetic field strength, as a function of input
parameters only (i.e. parameters that explicitely enter the governing
equations), and can therefore be used 
before any simulation is performed (as opposed to scaling laws involving measured quantities such as 
$\Ray_Q^\star$ and $\fohm$).

\subsection{Control parameters}
\label{control}
Only four non-dimensional parameters can be introduced in the governing
equations (\ref{eq_NS}-\ref{eq_div}).
In our formulation, these are the Ekman number $\Ek$, the
Prandtl number $\Pra$, the magnetic Prandtl number $\Pm$ and the Rayleigh
number $\Ray$ (see section~\ref{equations}).
According to the Buckingham $\pi$ theorem, any additional non-dimensional quantity, e.g. the Elsasser number
$\Lambda \equiv \Lo^2 \Pm / \Ek$, can therefore be expressed as a function of the above four
non-dimensional control parameters. The choice of non-dimensional parameters is however 
non-unique (for example, the Roberts number $\q = \kappa/\eta$ could be used instead of the
magnetic Prandtl number $\Pm = \nu/\eta$).

\cite{Stelzer13} opened the way to a predictive scaling by expressing
$\Nu-1$, $\Ro$ and ${\Lo}/{\fohm^{1/2}}$ as a function of $\Ray$  
instead of $\Ray_Q^\star$ (see their section~5). 
Their approach however is bound to fail for small values of $\Ray$ as all
these measured quantities obviously vanish below the onset of convection or
dynamo action.

Instead of using the Rayleigh number as control parameter, it is thus natural to
introduce the distance to an instability threshold.
We thus introduce $\Ray_c$ and $\Ray_d$, which respectively 
correspond to the onset of convection and dynamo action (see appendix~\ref{AppSeuil} for $\Ray_d$). 
The measured quantities $\Nu-1$ and $\Ro$ are expected to vanish at the onset
of convection $\Ray=\Ray_c$ and ${\Lo}$ at the onset of dynamo action $\Ray=\Ray_d$~.

\begin{figure}
\centerline{\includegraphics[width=0.9\textwidth]{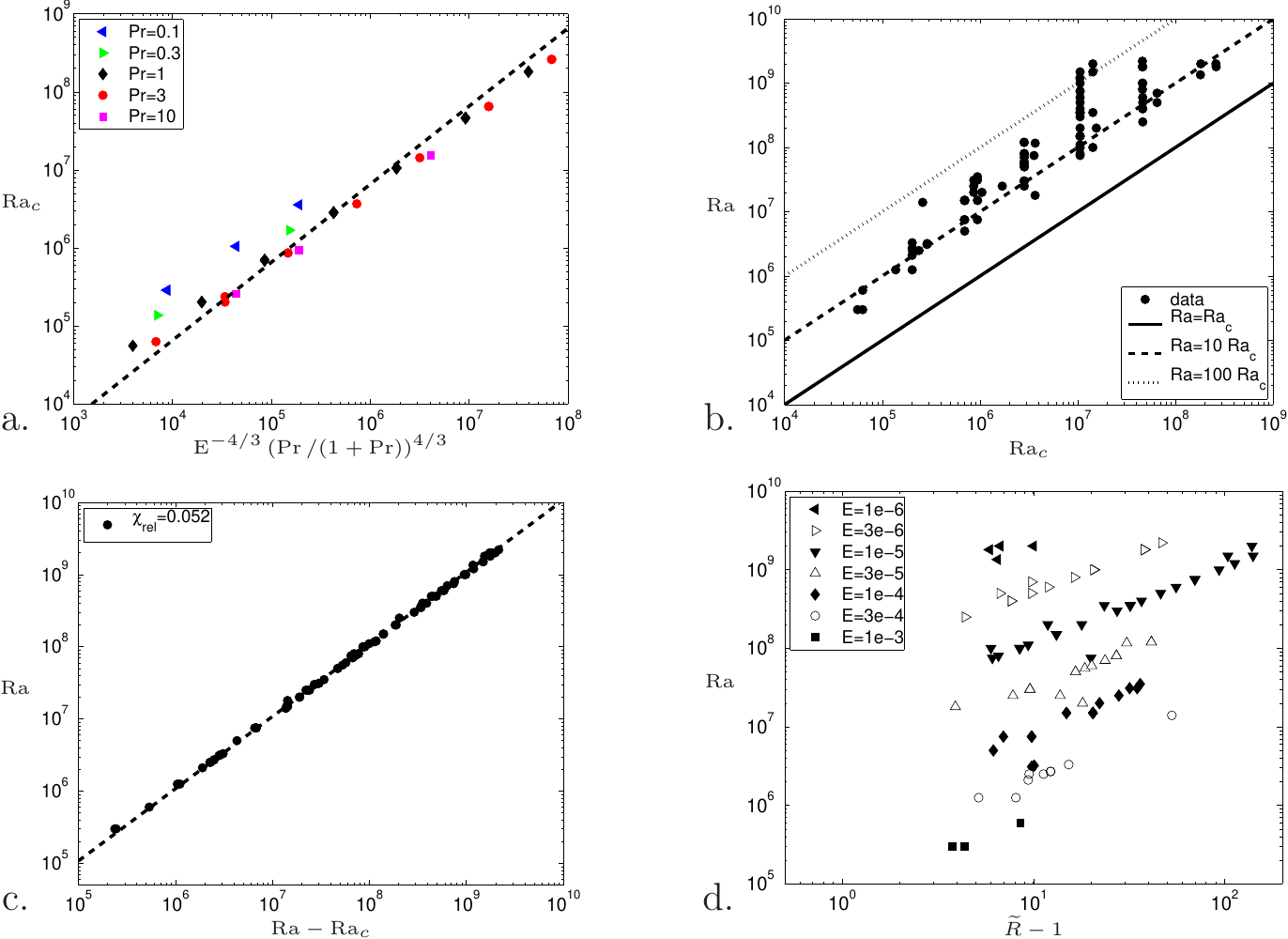}}
\caption{(a) The critical Rayleigh number for the onset of convection versus the predicted combination of the Ekman number 
and the Prandtl number 
\citep{Busse70}. (b) Parameter range: Rayleigh number in ordinate, 
critical Rayleigh number for convection in abscissa. 
(c) The strong correlation in the database between the Rayleigh number and its distance to the onset of convection. 
(d) The Rayleigh number versus its normalised distance to the onset of convection. The four graphs rely on the $102$ dynamos database.}
\label{Rac}
\end{figure}

The quantity $\Ray-\Ray_c$ therefore provides a natural control parameter
for hydrodynamic quantities such as $\Nu-1$ and $\Ro$.
This control parameter, even though natural, is however biased because of the strong
dependence of the critical Rayleigh number $\Ray_c$ on $\Ek$ and
$\Pra$. This dependence, first formulated and investigated by \cite{Chandra61} in the cartesian geometry, has been 
extensively studied. Especially, \cite{Roberts68} then \cite{Busse70} studied the limit $\Ek<\!<1$ in a spherical geometry. 
In a perturbative cylindric model for
a uniformly heated fluid, \cite{Busse70} proposed 
\begin{equation}
\Ray_c \sim \Ek^{-4/3}\, \left(\frac{\Pra}{1+\Pra}\right)^{4/3} \, .
\label{Rac_Busse}
\end{equation}
This solution, valid in the limit of asymptotic Ekman numbers, is consistent with several other studies: e.g. \cite{Carrigan83} 
(experimental convection study in a differentially heated 
spherical shell), \cite{Jones00} (uniformly heated fluid in a sphere), \cite{Takehiro02} 
(fixed heat flux boundary conditions), \cite{Dormy04} and \cite{Zhang04}.
It is validated to a certain extent against the finite Ekman number numerical database 
used in this paper. The corresponding misfit is $\chirel=0.319$ and it is represented in  
figure~\ref{Rac}.a. Note that a dependence on $\Pra/(1+\Pra)$ remains. The optimised scaling law 
obtained with our database is $\Ray_c \simeq 17.78 \, \Ek^{-1.19}\, (\Pra/(1+\Pra))^{0.58}$, with $\chirel=0.061$ 
($95\%$ confidence intervals in table~\ref{incertitudes_hydro}): optimised exponents are slightly
weaker (in absolute value) than those predicted by the asymptotic calculus of \cite{Busse70}, which indicates that these models
are still not in an asymptotic limit.

In practice, the numerical experiments used in this study are performed for
values of $\Ray$ of the order of $10$ times the critical value (see
figure~\ref{Rac}.b). Indeed, only dynamos with $\Nu>2$ are considered in the database \citep[see][]{Chr06}, 
on the other hand, for obvious computational
reasons associated with small scale motions, $\Ray$ is never very far from the
onset in numerical models. As a result, the values of $\Ray$ are strongly correlated with
the values of $\Ray_c$. It follows that \(\Ray\) is in fact close to
\(\Ray-\Ray_c\): in the numerical database, $\Ray \approx 1.11
(\Ray-\Ray_c)$ with a relative misfit $\chirel=0.052$ (see
figure~\ref{Rac}.c). This last relation, which traduces a bias in the database, explains why
\cite{Stelzer13} obtained satisfying fits of $\Ro$ and $\Nu-1$ as a
function of $\Ray$ 
(without introducing the distance to the onset of convection).  

The strong dependence of \(\Ray_c\) on the Ekman number introduces a very
large variation of the control parameter $\Ray-\Ray_c$, spanning over 
five orders of magnitude in the numerical database. This is somewhat
fictitious as the parameter \(\Ray/\Ray_c\) would only vary over one order of
magnitude. We thus introduce 
$\widetilde{R} \equiv {\Ray}/{\Ray_c}$ 
and our new control parameter will thus be
$\widetilde{R}-1$ (as $\widetilde{R}_c=1$). This new control parameter filters out the Ekman and Prandtl number 
dependences (the Ekman dependence is highlighted in figure~\ref{Rac}.d). 

{To measure the distance to the onset of dynamo action, 
we also introduce the control parameter $\widetilde{R}-\widetilde{R}_d$, where
$\widetilde{R}_d \equiv \Ray_d/\Ray_c$
is a function of $\Ek$, $\Pra$ and $\Pm$ only.
Nevertheless, whereas $\Ray_c$ is known for all numerical experiments in the database
provided by U.~Christensen, this is not the case for the critical value at
the onset of dynamo action $\Ray_d$.  It can be estimated through a linear interpolation of $\Lo^2$ as a function
of $\Ray$ near the onset of dynamo action (see appendix~\ref{AppSeuil}). 
Such an estimate could only be performed for seven sets of $\Ek$, $\Pra$
and $\Pm$ in the database (see table~\ref{dynamo_seuil}), which corresponds to 
$33$ numerical simulations. 
It is extended to $42$ simulations thanks to $9$ additional direct 
numerical simulations extracted from \cite{Morin09} and corresponding to
the set $\Ek=3 \times 10^{-4}$, $\Pra=1$, $\Pm=3$. 

Our four control parameters therefore are:  the Ekman number $\Ek$, the
Prandtl number $\Pra$, the magnetic Prandtl number $\Pm$ and the relative distance
to either the onset of convection or of dynamo action,
$\widetilde{R}-1$ and $\widetilde{R}-\widetilde{R}_d$ respectively.

\begin{table}
\centering
\begin{tabular}{c c c c c} 
\hline
$\Ek$ & $\Pra$ & $\Pm$ & $\Ray_d$ & $\Rm_d$\\
\hline
$3 \times 10^{-4}$ & $1$ & $3$ & $6.125 \times 10^{5}$ & $62.5$\\
$1 \times 10^{-4}$ & $1$ & $0.5$ & $3.6 \times 10^{6}$ & $26$\\
$1 \times 10^{-4}$ & $1$ & $1$ & $2.4 \times 10^{6}$ & $34$\\
$1 \times 10^{-4}$ & $10$ & $10$ & $5 \times 10^{5}$ & $25$\\
$3 \times 10^{-5}$ & $1$ & $0.25$ & $2.6 \times 10^{7}$ & $29$\\
$3 \times 10^{-5}$ & $1$ & $1$ & $1.5 \times 10^{7}$ & $70$\\
$3 \times 10^{-5}$ & $1$ & $2.5$ & $1.04 \times 10^{7}$ & $103$\\
$1 \times 10^{-5}$ & $1$ & $0.5$ & $8.0 \times 10^{7}$ & $70$\\
$1 \times 10^{-5}$ & $1$ & $1$ & $4.7 \times 10^{7}$ & $68$\\
$1 \times 10^{-5}$ & $1$ & $2$ & $4.9 \times 10^{7}$ & $150$\\
$3 \times 10^{-6}$ & $1$ & $0.1$ & $6.4 \times 10^{8}$ & $40$\\
$3 \times 10^{-6}$ & $1$ & $0.5$ & $2.4 \times 10^{8}$ & $60$\\
\hline
\end{tabular}
\caption{Estimated values of the Rayleigh number and of the magnetic Reynolds number, corresponding to the onset 
of dynamo action (see appendix~\ref{AppSeuil}).}
\label{dynamo_seuil}
\end{table}

\subsection{Direct numerical fit versus forces balances}
\label{limits}
Empirical scaling laws deduced from the multiple linear regression method
applied to numerical data have to be considered carefully for two main
reasons. First, the ranges of some input parameters are highly correlated,  
which introduces bias in scaling laws. It is the case for the Ekman number and the magnetic Prandtl number. 
The minimal value of $\Pm$ required for dynamo is indeed dependent on $\Ek$ \citep[see][]{Chr06}.
Figure~\ref{biais_fig} represents the range of $\Pm$ as a function of the range of $\Ek$ in the $102$ dymanos database used in the
present study: the minimal value
of $\Pm$ varies roughly as $\Ek^{3/4}$ \citep{Chr99,Chr06}, although this
cannot be distinguished from $\Ek^{2/3}$ \citep[as proposed
  by][]{Dormy08}. 
As a consequence, the scaling laws obtained via a direct numerical fit 
have to be considered carefully. In particular, biases can occur
relating dependences on $\Ek$ and $\Pm$. 

The second important limit of the approach based on empirical scaling laws deals with the restriction of our scaling analysis 
to power laws. In particular, the dependence on the Prandtl 
coefficient seems more complex than a simple power law. For instance, the dependence of the critical Rayleigh number 
$\Ray_c$ on $\Pra$ takes the form $\Pra/(1+\Pra)$ (see (\ref{Rac_Busse}) above). Indeed, a power law expression would diverge 
in the limit $\Pra$ tends to infinity.

\begin{figure}
\centerline{\includegraphics[width=0.42\textwidth]{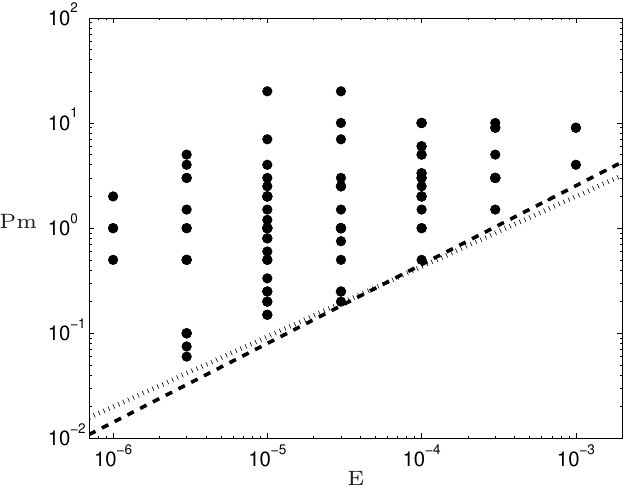}}
\caption{Correlation in control parameters used in numerical models; the magnetic Prandtl number is represented 
as a function of the Ekman number. The dashed line corresponds to $\Pm_{\rm min} = 450\,\Ek^{0.75}$ \citep{Chr06} and the dotted line 
to $\Pm_{\rm min} \sim \Ek^{2/3}$ \citep{Dormy08}. This figure relies on the full $102$ dynamos database.}
\label{biais_fig}
\end{figure}

Because of the above limitations, we prefer to guide our derivation of
scaling laws with physical arguments such as forces balances.
Our motivation is to take some distance with empirical fits,  and to
rely on the numerical database to validate the proposed scaling laws,
guided by physical arguments.   

\subsection{Magnetic field strength as a function of the flow amplitude}

A first step in our reasoning consists in expressing the magnetic field
strength as a function of the flow amplitude. 
In experimental physics, one usually controls the peak velocity of a flow
driven say by propellers. For this reason,  
earlier theoretical work often focused on the relation between the produced
magnetic field and the velocity field. 
A first approach is to consider dynamos which bifurcate from a laminar
flow. One assumes that in such cases, a dominant balance  
exists between the Lorentz force and the viscous force associated to the
flow modification \citep{Petrelis01}.  

It yields the equilibrium
\begin{equation}
\Lo ^2 \sim \Ek \, \frac{\Ro-\Ro_d}{{\tilde{\ell}_u}^{\star\, 2}} \, ,
\label{FP1}
\end{equation}
where $\Ro_d$ corresponds to the Rossby number at the onset of dynamo
action, and
the length scale $\tilde{\ell}_u$ corresponds to the 
characteristic length scale of the flow calculated as the mean scale
of the kinetic energy spectrum \citep[see][]{Chr06}. This length scale is very similar to our
$\ell_u$ introduced in appendix~\ref{section_fohm}.   

Supposing, as do \cite{Petrelis01}, that
${\tilde{\ell}_u}^{\star} \sim 1$, this leads in non-dimensional form to
\begin{equation}
{\Lambda} \sim {\left(\Rm-\Rm_d\right)} \, {\Ek} \, ,
\label{FP2}
\end{equation}
where $\Rm_d$ corresponds to the critical value of $\Rm$ at the onset of
dynamo action.

While the length scale ${\tilde{\ell}_u}^{\star}$ necessarily varies over a limited
range in the numerical database (see figure~\ref{fig_lu}.a and the
discussion at the end of section~\ref{inter_phys}), a finer
description can be achieved by retaining viscous effects and neglecting
inertial forces.
The equilibrium between the curl of the Coriolis force and 
the viscous force indeed yields
\begin{equation}
{\tilde{\ell}_u}^{\star} \sim \Ek^{1/3} \, .
\label{King2}
\end{equation}
This last relation properly describes the database used in the present paper, 
as shown in figure~\ref{fig_lu}.a and in \cite{King13} \citep[see also][]{RobertsKing13}. The lengthscale
${\tilde{\ell}_u}^{\star}$ clearly depends on $\Ek^{1/3}$ and not on  
$\Ek^{1/2}$, the latter being the typical scale of boundary layers. 
Thus, viscous effects play a non-negligible role in the bulk of the flow. 
This indicates that present numerical simulations are not in a
dynamical regime relevant to the Earth's core \citep[see also][]{Soderlund12}. The $\Ek^{1/3}$ scale would
represent less than 100m for geophysical values. Besides, the mild dependence 
of ${\tilde{\ell}_u}^{\star} \, \Ek^{-1/3}$ on the Rossby number 
(see figure~\ref{fig_lu}.b) shows that the assumption that inertia is small
compared to viscous effects is verified by numerical models.

\begin{figure}
\centerline{\includegraphics[width=0.9\textwidth]{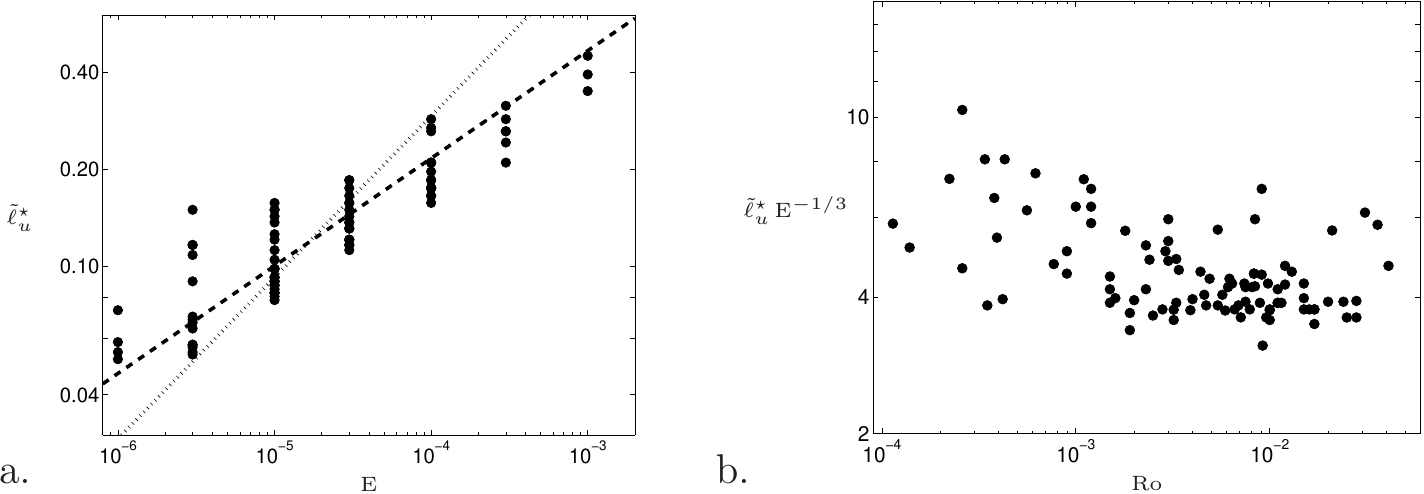}}
\caption{(a) The non-dimensional characteristic length scale ${\tilde{\ell}_u}^{\star}$ as a function of the 
Ekman number. The dashed line corresponds to ${\tilde{\ell}_u}^{\star} \sim \Ek^{1/3}$ (equation (\ref{King2}))
and the dotted line to ${\tilde{\ell}_u}^{\star} \sim \Ek^{1/2}$. (b) The corrected length scale 
${\tilde{\ell}_u}^{\star} \,\Ek^{-1/3}$ 
as a function of the Rossby number. Similar graphs can be produced using $\ell_u^{\star}$ instead of 
${\tilde{\ell}_u}^{\star}$. This figure relies on the full $102$ dynamos database.}
\label{fig_lu}
\end{figure}

If one uses (\ref{King2}) for the length scale ${\tilde{\ell}_u}^{\star}$
in relation (\ref{FP1}) \citep[see][]{Dormy07}, this yields
\begin{equation}
{\Lambda} \sim {\left(\Rm-\Rm_d\right)} \, {\Ek}^{1/3} \, .
\label{FP3}
\end{equation}
An alternative forces balance, known as the strong field balance, and assumed to be valid for the Earth's core, consists
in assuming a balance between the Lorentz force  
and the modification of the Coriolis force.
It provides \citep[see][]{Petrelis01} in non-dimensional form 
\begin{equation}
{\Lambda} \sim {\left(\Rm-\Rm_d\right)} \, .
\label{FP4}
\end{equation}
Each of the relations (\ref{FP2}), (\ref{FP3}) and (\ref{FP4}) can be tested against 
the $42$ dynamos database (see figure~\ref{Els_phys_fig1}), and
yields the relative misfits
$\chirel=1.438 \, , \ \chirel=0.891 \, ,$ and $\chirel=2.081$ respectively.
The best scaling law fitting the numerical data is therefore (\ref{FP3}). 
It is consistent with the fact that viscous effects have been shown to play
a non-negligible role in the bulk of the flow in numerical models.
\begin{figure}
\centerline{\includegraphics[width=1\textwidth]{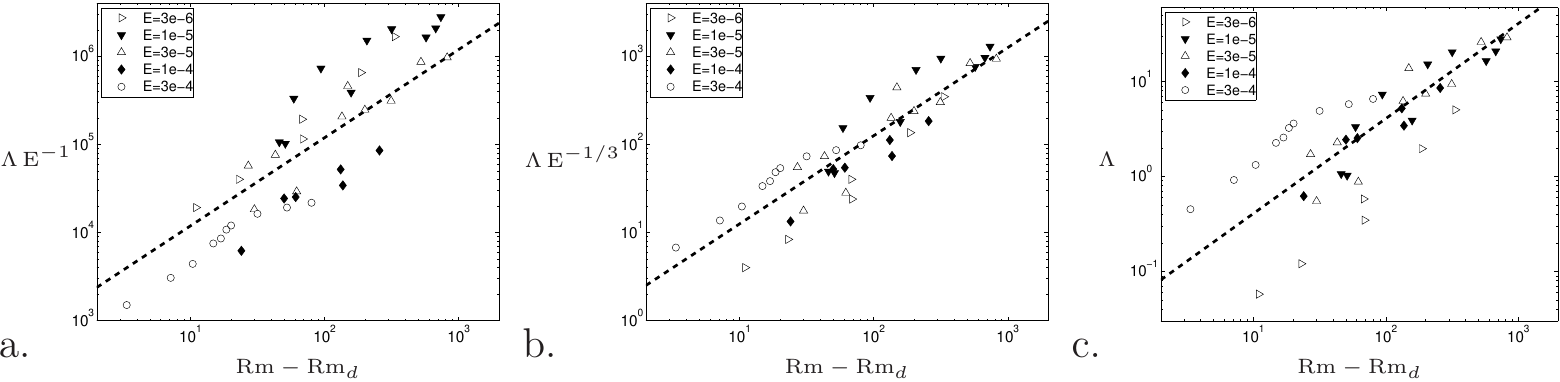}}
\caption{Scaling laws for the magnetic field strength as a function of the flow amplitude as measured by $\Rm-\Rm_d$: 
 (a) relation (\ref{FP2}), (b) relation (\ref{FP3}) and (c) relation (\ref{FP4}).
 This figure relies on the $42$ dynamos database.}
\label{Els_phys_fig1}
\end{figure}

This result can be compared to the output of a direct numerical fit. The values of
the magnetic Reynolds number at the onset of dynamo action corresponding to
seven sets of $\Ek$, $\Pra$ and $\Pm$ in the database have been estimated by a linear 
interpolation of $\Lo^2$ as a function of $\Rm$ (see table~\ref{dynamo_seuil} and appendix~\ref{AppSeuil}).
The multiple linear regression approach applied to the $42$ dynamos database provides the following 
scaling law for the Elsasser number $\Lambda$ as a function of 
$(\Rm-\Rm_d)$ and $\Ek$ ($95\%$ confidence intervals given in table~\ref{incertitudes_mag}): \footnote{
A direct numerical fit of $\Lambda$ as a function of $(\Rm-\Rm_d)$, $\Ek$, $\Pm$ and $\Pra$  yields 
${\Lambda} \simeq 0.30 \,{(\Rm-\Rm_d)}^{0.88} \,\Ek^{0.12} \, \Pm^{0.79}\,  \Pra^{-0.82}\, ,$ with $\chirel=0.301$
\label{Elsasser_Ro_best} (see table~\ref{incertitudes_mag} and figure~\ref{Elsasser_Rm_Elsasser_fig}.a). The proposed 
dependence on $\Pra$ is not strongly constrained, since the estimation of the optimal exponent of $\Pra$ is only based on $3$ simulations corresponding 
to $\Pra \ne 1$ (and for all three, $\Ek=1 \times 10^{-4}$, $\Pra=10$, $\Pm=10$, see table~\ref{dynamo_seuil}).
The proposed dependence is therefore clearly not robust. 
The bias between $\Ek$ and $\Pm$ in the database probably accounts for the smaller exponent of $\Ek$ and the extra dependence on $\Pm$
in the above relation compared to (\ref{Elsasser_Ro_best2}).}
\begin{equation}
{\Lambda} \simeq 10.24 \,{(\Rm-\Rm_d)}^{1.09} \,\Ek^{0.52} \, ,
\qquad 
\mbox{with}\quad
\chirel=0.698\, .
\label{Elsasser_Ro_best2}
\end{equation}
The physically derived scaling law (\ref{FP3}) is consistent with the
empirical scaling law (\ref{Elsasser_Ro_best2}) for the dependence on 
\((\Rm-\Rm_d)\). The optimal exponent of $\Ek$ is larger than the $1/3$ value predicted by (\ref{FP3}), 
and remains to be investigated.

\begin{figure}
\centerline{\includegraphics[width=1\textwidth]{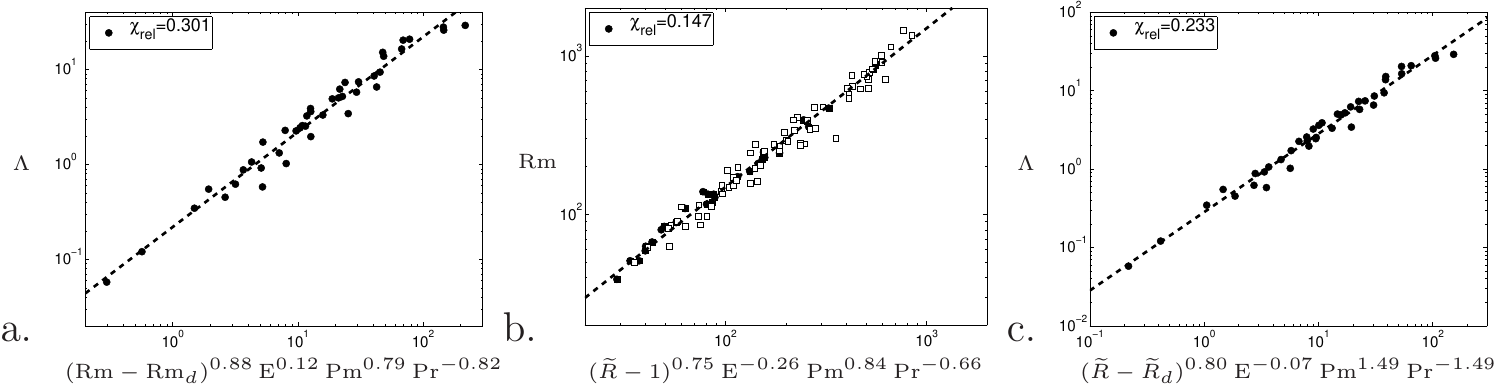}}
\caption{Scaling laws provided by direct numerical fits.
(a) The Elsasser number versus the flow amplitude (see footnote~\ref{Elsasser_Ro_best}).
(b) The magnetic Reynolds number versus the normalised distance to the
  onset of convection (see footnote~\ref{scaling_Rm}). 
(c) Predictive scaling law for the Elsasser number versus the normalised distance to the onset of dynamo action
(relation (\ref{Lo_dyn_3})).  Squares correspond to the $102$ dynamos database, 
black points to the $42$ runs of the reduced database.}
\label{Elsasser_Rm_Elsasser_fig}
\end{figure}

Figure~\ref{Els_phys_fig2}.a 
represents relation (\ref{FP3}) applied to the $42$ dynamos database in red diamonds, and the same relation, but setting $\Rm_d$ 
to zero in blue squares. The blue points gradually move away from a linear fit
when $\Rm$ decreases, as expected (because the approximation $\Rm-\Rm_d \simeq \Rm$ worsens).
Relation (\ref{FP3}) can however then be applied
to the $102$ dynamos database, provided that the parameter $\Rm_d$ is dropped 
(since it is only known for the $42$ simulations of the reduced database). It is represented in figure~\ref{Els_phys_fig2}.b. 
As in figure~\ref{Els_phys_fig2}.a, the full numerical database appears to follow the proposed scaling law, except for 
low values of $\Rm$ for which $\Rm_d$ cannot be neglected.

{It is worth noting that} {relation (\ref{FP3}) reveals a dependence of the magnetic field strength on viscosity, 
which is geophysically not realistic. To illustrate this, let us now try to
apply this relation to the Earth's core.} 
We choose the common estimate value $\Rm=10^3$.
The distance to the onset of dynamo action $\Rm-\Rm_d$ can be estimated by $\Rm$,
which leads to an over-estimated value of $\Lambda$.
We find $\Lambda \sim 10^{-2}$, which is an upper bound because $\Rm_d$ was not taken into
account. It is yet much smaller than its estimated value for the Earth's core, expected to be close to unity \citep{Roberts88}.
{This indicates very clearly that available numerical models
  are not in the dynamical regime relevant to geodynamo.} 
{In other words, the Earth's core would simply be out of
  the range of figure~\ref{Els_phys_fig2}.a (with $\Rm \simeq 10^3$, and
  $\Lambda \, \Ek^{1/3}\simeq 10^5$). } 

The magnetic Reynolds number is however a measured quantity in the numerical database. In order to establish
a predictive scaling for the magnetic field strength, it is thus necessary to express the flow amplitude as a function of control parameters. 
This is the purpose of the two next sections.

\begin{figure}
\centerline{\includegraphics[width=0.9\textwidth]{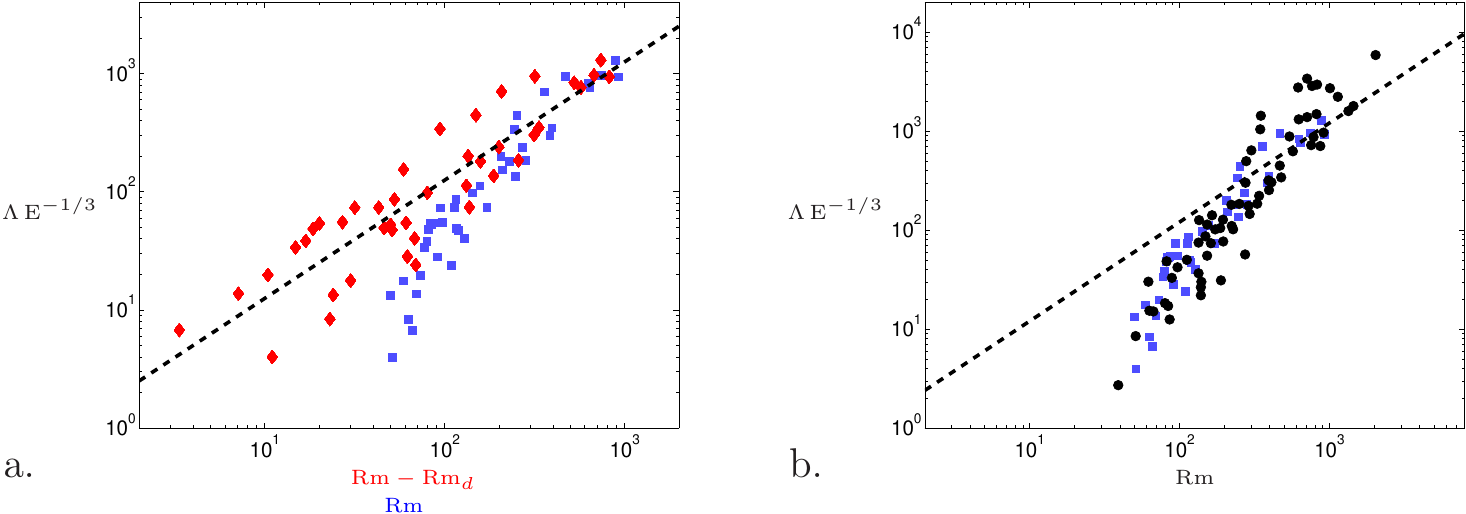}}
\caption{Physically derived relation for the magnetic field strength as a function of $\Rm-\Rm_d$.
(a) Relation (\ref{FP3}) (red diamonds), 
and the same relation but setting $\Rm_d$  to zero (blue squares), 
both applied to the $42$ dynamos database. (b) Relation (\ref{FP3}) dropping the unknown 
$\Rm_d$ contribution, applied to the full (blacks points) and
reduced (blue squares) database. The dashed line corresponds to relation (\ref{FP3}).} 
\label{Els_phys_fig2}
\end{figure}

\subsection{Predictive scaling law for the injected power}

{The definition of the output parameter ${\Ray^{\star}_Q}$ involves the efficiency with which 
heat is transferred by convection, measured by the Nusselt number $\Nu$ (see equation (\ref{def_RastarQ})). 
This is a subject of study in itself, many studies of heat transfer have been performed for rotating convection. 
Figure~\ref{Nu_fig} shows that the numerical data globally correspond to an intermediate regime between the rapidly rotating regime 
\citep[$\Nu=\widetilde{R}^{\,6/5}$,][]{Aurnou07,King09,King10} and the weakly rotating regime \citep[$\Nu \sim \Ray^{\,2/7}$, see][]{King09}. 
The simple relation} {$\Nu \sim \widetilde{R}$} {provides a good description of the database. Note that in figure~\ref{Nu_fig}, 
a dependence on $\Pra$ remains. This could be further investigated 
by seeking for a dependence on $\Pra/(1+\Pra)$ (instead of a power 
law dependence which would lack regularity in the limit $\Pra \longrightarrow 0$ or $\Pra \longrightarrow +\infty$). 
As we will however discuss later (see section~\ref{section_finale}), the
$\Pra/(1+\Pra)$ term can be omitted without significant loss of quality in
describing the present database.}

{In the available database, the bias $\Ray \sim \Ray-\Ray_c$ (see section~\ref{control}) allows us to approximate 
$\Ray$ by $\left(\widetilde{R}-1\right)\, \Ray_c$ in the definition (\ref{def_RastarQ}) of $\Ray^{\star}_Q$. 
Injecting (\ref{Rac_Busse}) and the above simple expression for $\Nu$ (which is admittedly not 
based on solid physical considerations) yields}
\begin{equation}
\Ray^{\star}_Q \sim (\widetilde{R}-1)^2 \, {\Ek}^{5/3}\, {\Pra}^{-2} \, \left(\frac{\Pra}{1+\Pra}\right)^{4/3} \,  .
\label{RastarQ_phys}
\end{equation}
{The corresponding relative misfit when applied to the $102$ dynamos database is $\chirel=0.311$ and it is 
represented in figure~\ref{RastarQ_fig}.a. 
Note that the dependence on $\Pra^{-2}$ in (\ref{RastarQ_phys}) 
comes from the definition (\ref{def_RastarQ}) of $\Ray^{\star}_Q$, whereas that on $(\Pra/(1+\Pra))^{4/3}$
results from the expression of $\Ray_c$ (relation (\ref{Rac_Busse})). }

Relation (\ref{RastarQ_phys}) can be compared to an optimised empirical fit, which provides}
\begin{equation}
\Ray_Q^\star \simeq 5.10 \,{(\widetilde{R}-1)}^{1.78}\, \Ek^{1.70}\, \Pra^{-2.12}\, \left(\frac{\Pra}{1+\Pra}\right)^{1.26}\, ,
\qquad \mbox{with}\quad \chirel=0.173 \, ,
\label{eqRastarQ3}
\end{equation}
($95\%$ confidence intervals are given in table~\ref{incertitudes_hydro}). 
Relation (\ref{RastarQ_phys}) is therefore close to providing the 
best fit through the numerical database.\footnote{
If the dependence on $\Pra/\left(1+\Pra\right)$ is omitted, it yields a larger misfit ($\chirel=0.326$), and
$\Ray_Q^\star \simeq 1.47\,{(\widetilde{R}-1)}^{1.77}\, \Ek^{1.68}\, \Pra^{-1.56}\, .$
\label{eqRastarQ2}
Both relations would not be modified if a dependence on $\Pm$ was sought (see table~\ref{incertitudes_hydro}).}

The parameter $\Ray_Q^\star$
varies over six orders of magnitude, while none of the control parameters
varies over such a wide range (see appendix~\ref{range_ra_star_Q}).
Figure~\ref{RastarQ_fig}.b highlights the strong dependence 
of $\Ray_Q^\star$ on the Ekman number explicited in
(\ref{RastarQ_phys}). This explains the above range of variation.
The control parameter \(\widetilde{R}-1\) covers a much more physically realistic range.

\begin{figure}
\centerline{\includegraphics[width=0.42\textwidth]{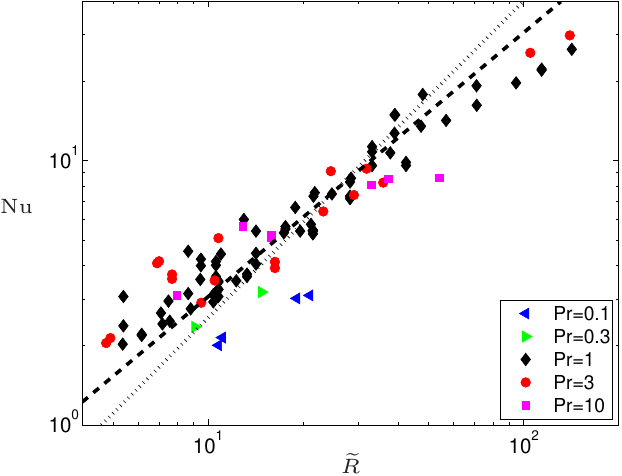}}
\caption{{The Nusselt number versus the Rayleigh number normalised by its critical value. 
The dotted line corresponds to $\Nu \sim \widetilde{R}^{\,6/5}$ and the dashed line to $\Nu \sim \widetilde{R}$. 
The Prandtl number is indicated by the color (as in figure~\ref{Rac}.a). This figure is based on the $102$ dynamos database.}}
\label{Nu_fig}
\end{figure}

\begin{figure}
\centerline{\includegraphics[width=0.9\textwidth]{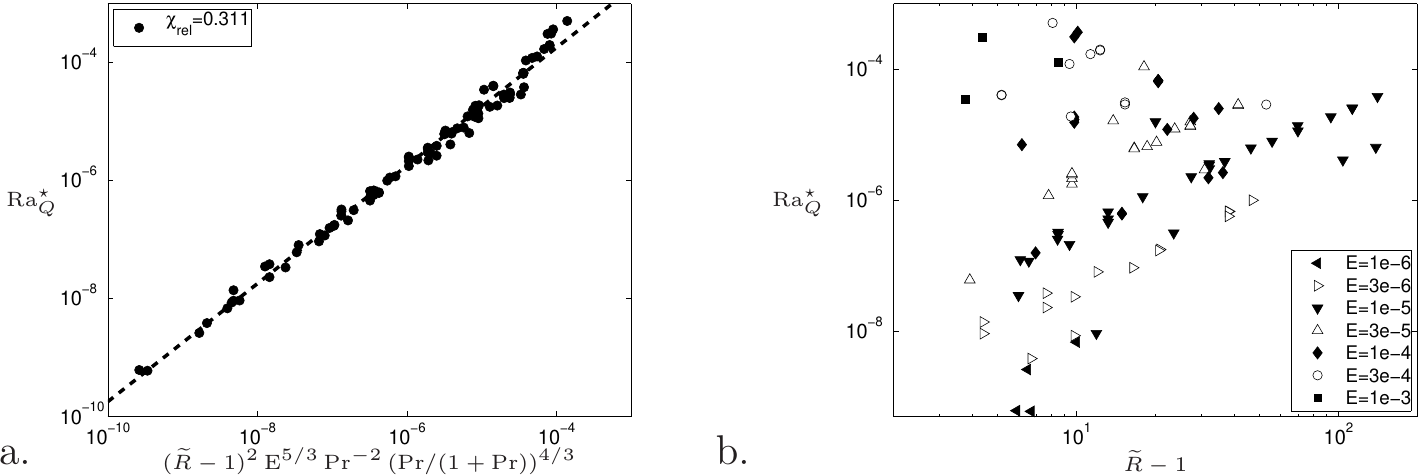}}
\caption{(a) Validation of relation (\ref{RastarQ_phys}) expressing the flux-based Rayleigh number 
as a function of the normalised distance to the onset of dynamo. (b) The flux-based Rayleigh number 
versus the normalised distance to the onset of convection. 
The Ekman number is indicated by using different symbols (as in figure~\ref{Rac}.d). 
This figure is based on the full $102$ dynamos database.}
\label{RastarQ_fig}
\end{figure}

\subsection{Predictive scaling laws for the flow amplitude}
\label{Ro_section}

\begin{figure}
\centerline{\includegraphics[width=0.9\textwidth]{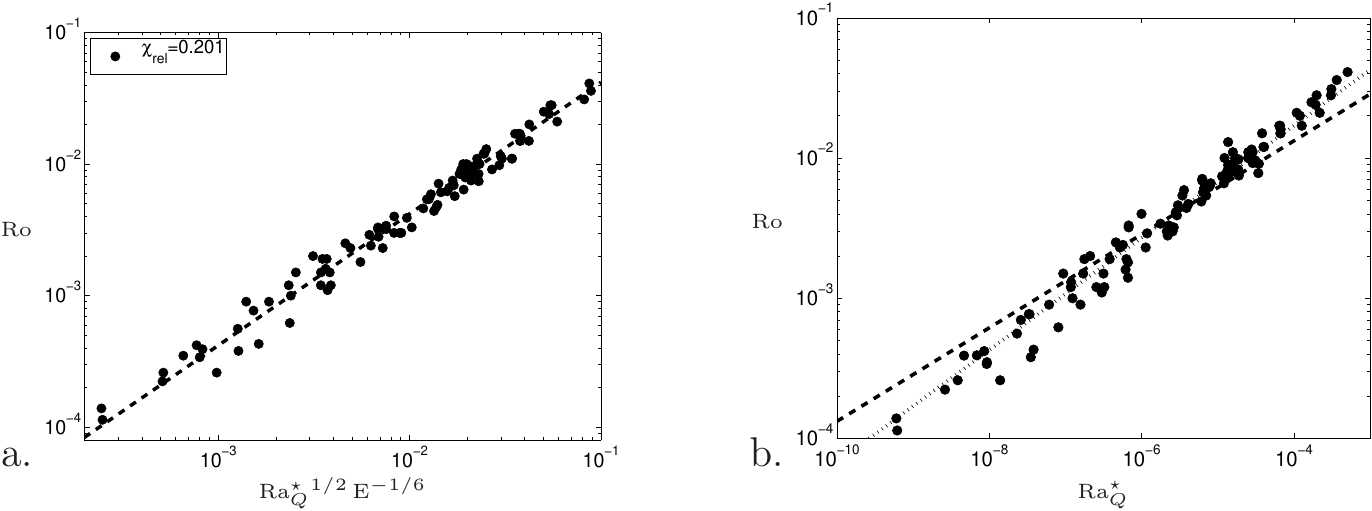}}
\caption{(a) The Rossby number as a function of the flux-based Rayleigh number (a) 
based on the VAC scaling (relation (\ref{VAC}))
(b) on the IAC scaling (dotted line, $\chirel=0.237$) and on the scaling resulting from mixing length theory 
(dashed line, $\chirel=0.431$). 
Both graphs rely on the full $102$ dynamos database.}
\label{Rm_phys_fig}
\end{figure}

Several scaling laws based on different forces balances have been
proposed in the literature concerning the flow amplitude \citep[detailed in][]{King13}. For simulations near the onset of dynamo action,
the Lorentz force can be expected to be small. Then, balancing the curl of the buoyancy term with that
of the Coriolis force yields 
\begin{equation}
u^{\star} \sim {\tilde{\ell}_u}^{\star \, -1}\, \frac{\Power^{\star}}{u^{\star}} \, ,
\label{King1}
\end{equation}
which can be rewritten, using relation (\ref{Power_CA06}), as
\begin{equation}
\Ro \sim \tildel_u^{\star \,-1/2 } \, {\Ray^{\star}_Q}^{1/2}\, .
\label{King1bis}
\end{equation}
Combining (\ref{King1bis}) and (\ref{King2}) leads to the Viscous-Archimedean-Coriolis (VAC) scaling proposed 
by \cite{King13} 
\begin{equation}
\Ro \sim {\Ray^{\star}_Q}^{1/2} \, \Ek^{-1/6} \, . 
\label{VAC}
\end{equation}
{Its application to the 
$102$ dynamos database is represented in figure~\ref{Rm_phys_fig}.a, with a relative misfit $\chirel=0.201$. 
It can be compared to the inertial $\Ro$-scalings $\Ro \sim {\Ray^{\star}_Q}^{2/5}$ \citep[derived from the IAC balance, see][]{Aubert01,Jones11} and $\Ro \sim {\Ray^{\star}_Q}^{1/3}$ \citep[resulting from mixing length theory, see][]{Chr10} 
(see figure \ref{Rm_phys_fig}.b). These three scaling laws provide descriptions of the database of comparable quality. 
The scaling law (\ref{King2}) for the length scale ${\tilde{\ell}_u}^{\star}$ however indicates that the VAC scaling (\ref{VAC}) 
is more relevant to the present study than inertial scaling laws. }

The direct multiple linear approach provides the following optimal scaling 
law expressing $\Ro$ as a function of $\Ray^{\star}_Q$ and $\Ek$
($95\%$ confidence intervals given in table~\ref{incertitudes_hydro}) \footnote{
A direct numerical fit for $\Ro$ as a function of $\Ray^{\star}_Q$, $\Ek$, $\Pm$ and $\Pra$ yields
\( 
\Ro \simeq  1.10\, {\Ray^{\star}_Q}^{0.43} \, \Pm^{-0.14}, \)
with
\(\chirel=0.100 \)
\label{scaling_Ro2}
($95\%$ confidence intervals given in table~\ref{incertitudes_hydro}). 
The role of $\Ek$ and $\Pra$ are found to be negligible. However the bias in the database (see section~\ref{limits}) 
renders the dependence on $\Pm$ unreliable (as $\Ek$ and $\Pm$ are
correlated in the database).}~:
\begin{equation}
\Ro \simeq  0.59\, {\Ray^{\star}_Q}^{0.47}\, {\Ek}^{-0.10}, 
\qquad
\mbox{with}\quad
\chirel=0.184\, .
\label{scaling_Ro}
\end{equation}

\begin{table}
\scriptsize
\centering
\begin{tabular}{c c c c c c c c c} 
\hline\\[-2.5mm]
 &  Pre-factor & $\Ray^{\star}_Q$ & $\widetilde{R}-1$ & $\Ek$ & $\Pm$ & $\Pra$ & $\Pra/(1+\Pra)$ & $\chirel$  \\
\hline
$\Ray_c$ & $17.779 \pm 1.468$ & $\times$ & $\times$ & $-1.193 \pm 0.008$ & $\times$ & $\times$ & $0.579 \pm 0.030$ & $0.061$ \\
$\Ray_Q^{\star}$ &  $1.470 \pm 0.517$  & $\times$ & $1.774 \pm 0.064$ & $1.675 \pm 0.032$ & - & $-1.557 \pm 0.061$ & $\times$ & $0.326$ \\
$\Ray_Q^{\star}$ &  $5.103 \pm 1.550$  & $\times$ & $1.775 \pm 0.041$ & $1.703 \pm 0.021$ & - & $-2.124 \pm 0.102$ & $1.256 \pm 0.208$ & $0.173$ \\
$\Ro$ & $0.589 \pm 0.133$ & $0.466 \pm 0.018$ & $\times$ & $-0.095 \pm 0.033$ & $\times$ & $\times$ & $\times$ & $0.184$\\
$\Ro$ & $1.103 \pm 0.094$ & $0.433 \pm 0.006$ & $\times$ & - & $-0.137 \pm 0.015$ & - & $\times$ & $0.100$\\
$\Rm$ & $ 1.535\pm 0.371$ & $\times$ & $ 0.749\pm 0.036 $ & $ -0.264\pm 0.020$ & $ 0.843\pm 0.030$ & $-0.656\pm 0.035$ & $\times$  & $0.147$\\
$\Rm$ & $2.421 \pm 0.547$ & $\times$ & $0.757 \pm 0.029$ & $-0.257 \pm 0.016$ & $0.857 \pm 0.024$ & $-0.901 \pm 0.070$ & $0.528 \pm 0.139$ & $0.108$\\
\hline
\end{tabular}
\caption{Optimal scaling laws obtained by the multiple linear regression method, for $\Ray_c$, $\Ray_Q^{\star}$ 
(relation (\ref{eqRastarQ3}) and relation given in the footnote~\ref{eqRastarQ2}), $\Ro$ (relation
  (\ref{scaling_Ro}) and relation given in the footnote~\ref{scaling_Ro2})
and $\Rm$ (relation given in the footnote~\ref{scaling_Rm} and
relation (\ref{scaling_Rm_opt})) ($95\%$ confidence intervals). Crosses indicate that
the corresponding parameter is chosen not to enter the fit. The dashes indicate that the contribution of  
the corresponding parameter has been found negligible.}
\label{incertitudes_hydro}
\normalsize
\end{table}

\begin{figure}
\centerline{\includegraphics[width=0.42\textwidth]{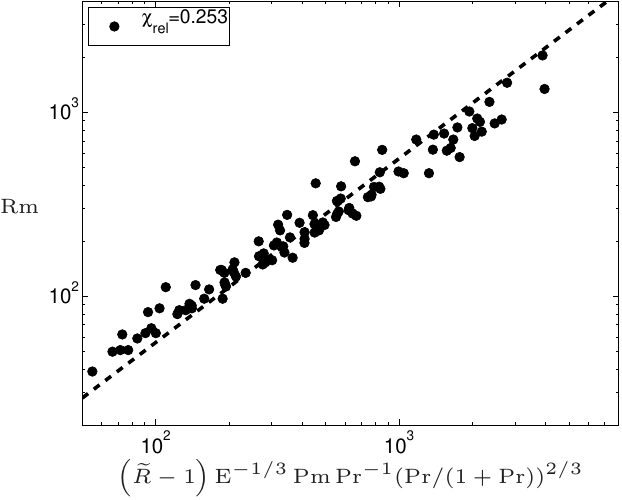}}
\caption{The resulting predictive scaling law for the magnetic Reynolds number 
as a function of the normalised distance to the onset of convection (relation (\ref{Rm_phys})). 
This figure relies on the full $102$ dynamos database.}
\label{Rm_phys_fig2}
\end{figure}

\begin{figure}
\centerline{\includegraphics[width=0.9\textwidth]{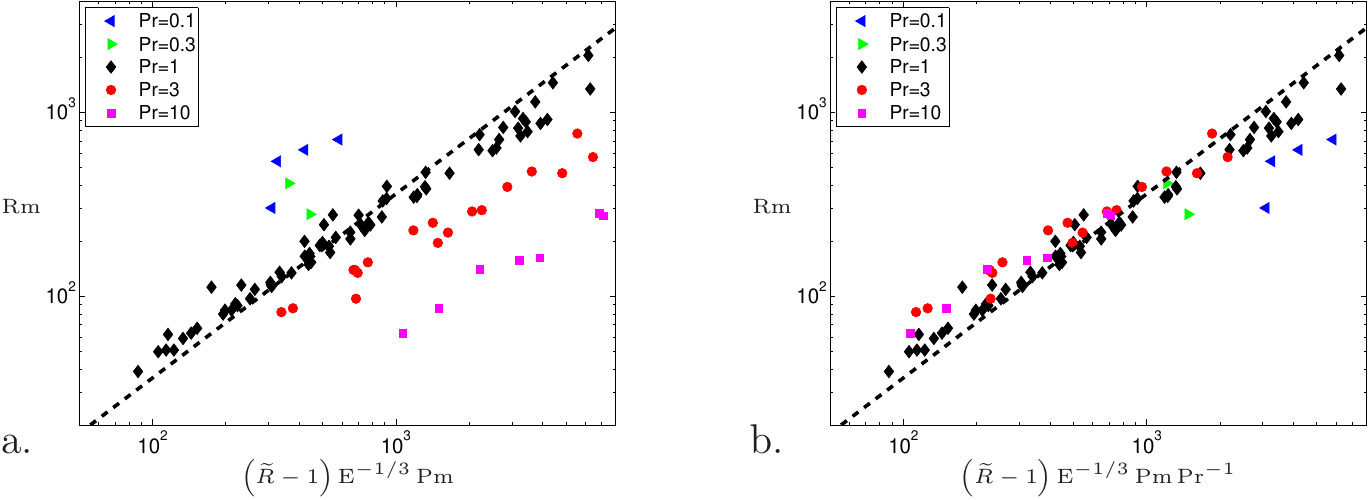}}
\caption{Simplified expressions for $\Rm$ testing the $\Pra$ dependence. 
The magnetic Reynolds number as a function of the normalised distance to the onset of convection, 
equation (\ref{Rm_phys}) (a) with no correction on $\Pra$ and $\Pra/(1+\Pra)$ (b) with the correction on $\Pra^{-1}$ only. 
Both graphs rely on the $102$ dynamos database.}
\label{Rm_phys_fig_rolePr}
\end{figure}

\begin{table}
\scriptsize
\centering
\begin{tabular}{c c c c c c c c} 
\hline\\[-2.5mm]
 &  Pre-factor & $\Rm-\Rm_d$ & $\widetilde{R}-\widetilde{R}_d$ & $\Ek$ & $\Pm$ & $\Pra$ & $\chirel$  \\
\hline
$\Lambda$ & $10.243 \pm 8.619$ & $1.091 \pm 0.157$ & $\times$ & $0.516 \pm 0.132$ & $\times$ & $\times$ & $0.698$\\
$\Lambda$ & $0.305 \pm 0.212$ & $0.879 \pm 0.087$ & $\times$ & $0.119 \pm 0.101$ & $0.787 \pm 0.141$ & $-0.820 \pm 0.174$ & $0.301$\\
$\Lambda$ & $0.351 \pm 0.210$ & $\times$ & $0.796 \pm 0.062$ & $-0.072 \pm 0.071$ & $1.490 \pm 0.096$ & $-1.491 \pm 0.144$ & $0.233$ \\
\hline
\end{tabular}
\caption{Optimal scaling laws obtained by the multiple linear regression method, for $\Lambda$ as a 
function of $(\Rm-\Rm_d)$ (relation (\ref{Elsasser_Ro_best2}) and relation given in footnote~\ref{Elsasser_Ro_best}) and $(\widetilde{R}-\widetilde{R}_d)$ (relation (\ref{Lo_dyn_3})) 
($95\%$ confidence intervals).}
\label{incertitudes_mag}
\normalsize
\end{table}
Replacing the parameter $\Ray^{\star}_Q$ by its expression (\ref{RastarQ_phys}) in equation (\ref{VAC}) yields
\begin{equation}
\Rm \sim (\widetilde{R}-1) \, {\Ek}^{-1/3}\, {\Pm} \, {\Pra}^{-1} \, \left(\frac{\Pra}{1+\Pra}\right)^{2/3} \,  . 
\label{Rm_phys}
\end{equation}
{This relation is essential in order to establish a predictive scaling law for the magnetic field strength, 
whereas relations (\ref{RastarQ_phys}) and (\ref{VAC}) are only intermediate steps in the reasoning. Besides, the dependence of 
$\Rm$ on ${\Ek}^{-1/3}$ counterbalances the dependence of $\Lambda$ on ${\Ek}^{1/3}$ in (\ref{FP3}), and 
thus removes the dependence of $\Lambda$ on viscosity in its predictive form.
The scaling law (\ref{Rm_phys})} applied to the $102$ dynamos database is represented in figure~\ref{Rm_phys_fig2}, 
with a relative misfit $\chirel=0.253$. 
The role of both terms ${\Pra}^{-1}$ and $\left(\Pra/(1+\Pra)\right)^{2/3}$ is illustrated in figure~\ref{Rm_phys_fig_rolePr}: 
the term $\Pra/(1+\Pra)^{2/3}$ allows to correct the data corresponding to
weak values of $\Pra$. 
{The non-negligible dependence on
$\Pra/\left(1+\Pra\right)$  is consistent with previous studies of
convection which established a dependence of the velocity amplitude on
$\Pra$ more complex than a simple power law dependence \citep[e.g.][]{Schluter65,Tilgner96}.}

Finally, note that relation (\ref{Rm_phys}) can be compared
to the optimised fit to the available data.
A direct numerical fit provides \footnote{
Omitting the dependence on $\Pra/\left(1+\Pra\right)$ provides a larger misfit (\(\chirel=0.147\)) and
\(
\Rm \simeq 1.54 \, \left(\widetilde{R}-1\right)^{0.75} \Ek^{-0.26} \Pm^{0.84}
\Pra^{-0.66}\)
\label{scaling_Rm}
(see figure~\ref{Elsasser_Rm_Elsasser_fig}.b and table~\ref{incertitudes_hydro}).}
\begin{equation}
\Rm \simeq 2.42 \, \left(\widetilde{R}-1\right)^{0.76} \Ek^{-0.26} \Pm^{0.86} \Pra^{-0.90} \left(\frac{\Pra}{1+\Pra}\right)^{0.53} \, ,\qquad
\mbox{with}\quad
\chirel=0.108\, ,
\label{scaling_Rm_opt}
\end{equation}
(the $95\%$ confidence intervals 
are provided in table~\ref{incertitudes_hydro}).
The exponents in relations (\ref{Rm_phys}) and (\ref{scaling_Rm_opt}) match to within $20\%$. 

\begin{figure}
\centerline{\includegraphics[width=0.9\textwidth]{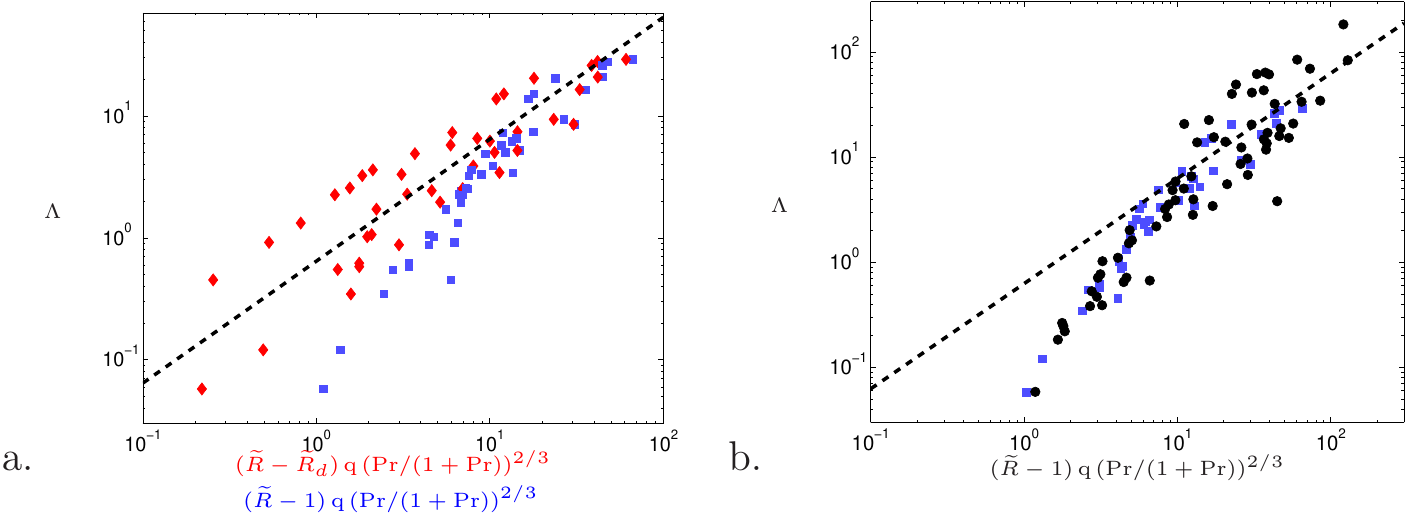}}
\caption{Physically derived predictive scaling law for the magnetic field
  strength as a function of $\widetilde{R}-\widetilde{R}_d$. 
(a) Relation (\ref{Els_phys_Rtild}) (red diamonds), 
and the same relation but approximating $\widetilde{R}_d$ to unity (blue squares), 
both applied to the $42$ dynamos database. (b) Relation (\ref{Els_phys_Rtild}) approximating the unknown 
$\widetilde{R}_d$ contribution to unity, applied to the full (blacks points) and
reduced (blue squares) database. The dashed line corresponds to relation (\ref{Els_phys_Rtild}).}
\label{Els_phys_fig3}
\end{figure}

\subsection{Predictive scaling law for the magnetic field strength}
\label{section_finale}
Replacing the flow amplitude in relation (\ref{FP3}) by its expression (\ref{Rm_phys}) yields the following 
predictive scaling law
\begin{equation}
\Lambda \sim (\widetilde{R}-\widetilde{R}_d) \, {\Pm} \, {\Pra}^{-1} \,
\left(\frac{\Pra}{1+\Pra}\right)^{2/3} 
\,  .
\label{Els_phys_Rtild}
\end{equation}
The direct numerical fit provides in the form of pure power laws
\begin{equation}
\Lambda \simeq 0.35 \,{(\widetilde{R}-\widetilde{R}_d)}^{0.80} \,\Ek^{-0.07} \, \Pm^{1.49}\,  \Pra^{-1.49}\, ,
\qquad 
\mbox{with}\quad
\chirel=0.233\, ,
\label{Lo_dyn_3}
\end{equation}
(see figure~\ref{Elsasser_Rm_Elsasser_fig}.c and table~\ref{incertitudes_mag} for $95\%$ confidence intervals). 
The dependence on the Ekman number is here negligible. Besides, we used here the reduced $42$ 
dynamos database for which $\widetilde{R}_d$ can be estimated, the coefficients based on a direct numerical fit 
are therefore weakly constrained. In particular $\Pra$ does not vary much
in this subsample.
Despite of this, the agreement between both expressions is remarkably good, except for a
larger exponent of $\Pm$ for the latter, which remains to be investigated.

The application of relation (\ref{Els_phys_Rtild}) to the $42$ dynamos database is
represented in figure~\ref{Els_phys_fig3}.a in red diamonds. 
The same expression approximating 
$\widetilde{R}_d$ to unity is plotted using blue squares. As expected, the quality of the approximation decreases with $\widetilde{R}$. 
Finally, figure~\ref{Els_phys_fig3}.b corresponds 
to the application of relation (\ref{Els_phys_Rtild}) to the full database, approximating 
the unknown $\widetilde{R}_d$ contribution to unity.

\begin{figure}
\centerline{\includegraphics[width=0.9\textwidth]{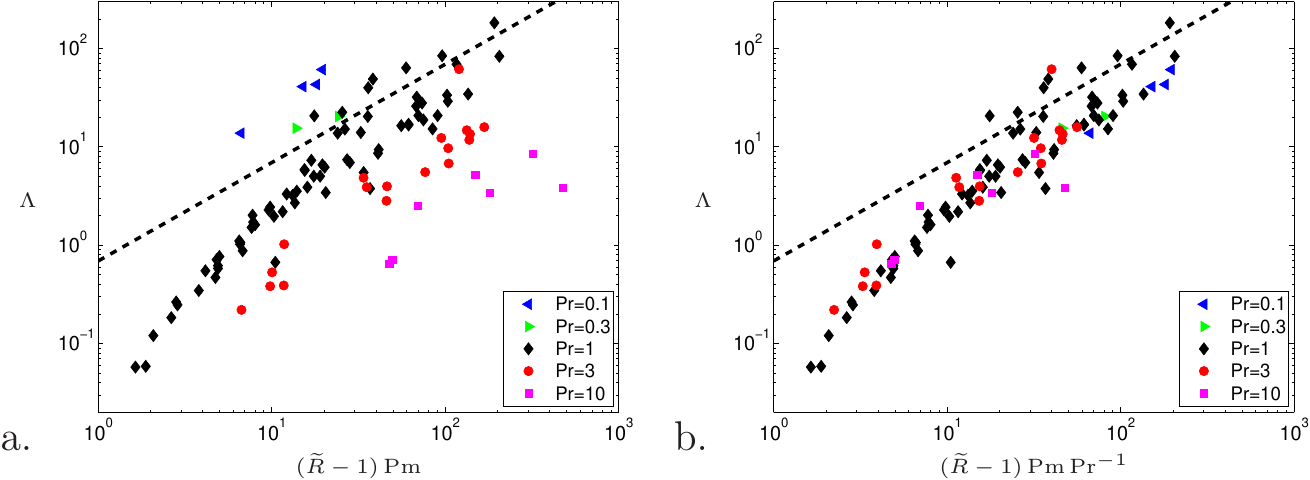}}
\caption{Test of the $\Pra$ dependence in the predictive scaling law (\ref{Els_phys_Rtild}) applied 
to the $102$ dynamos database approximating the unknown $\widetilde{R}_d$ contribution to unity 
(a) with no correction on $\Pra$ and $\Pra/(1+\Pra)$ (b) with the correction on $\Pra^{-1}$ only.
The dashed line in (a) and (b) corresponds to the scaling law (\ref{Els_phys_Rtild}).} 
\label{Els_phys_fig_rolePr}
\end{figure}

Finally, in order to assess the role of the two terms ${\Pra}^{-1}$ and $\left(\Pra/(1+\Pra)\right)^{2/3}$ in relation (\ref{Els_phys_Rtild}), 
we compare in figure~\ref{Els_phys_fig_rolePr} the improvements obtained by each contribution of $\Pra$. It highlights the important role 
of the ${\Pra}^{-1}$ term. The role of the $\Pra/(1+\Pra)$ term 
in the description of the available numerical database is marginal (compare
figures~\ref{Els_phys_fig3}.b and \ref{Els_phys_fig_rolePr}.b).

Thus, instead of the power based scaling law proposed by \cite{Chr10}, which can be rewritten as 
\begin{equation}
{\Lambda} \sim {\fohm}\,{\Ray_Q^\star}^{2/3} \,\Ek^{-1}\, \Pm \, ,
\qquad 
\mbox{with}\quad
\chirel=0.452\, ,
\label{fitCA06_nonopt_mod}
\end{equation}
and which involves measured quantities ($\fohm$ and $\Ray_Q^\star$), we
propose the simple relation (\ref{Els_phys_Rtild}), which can be
reformulated as
\begin{equation}
{\Lambda} \sim (\widetilde{R}-\widetilde{R}_d) \, \q \left(\frac{\Pra}{1+\Pra}\right)^{2/3}\, ,
\qquad 
\mbox{with}\quad
\chirel=0.516\, .
\label{notrefit}
\end{equation}
The $\Pra/(1+\Pra)$ dependence comes from the asymptotic expression of the critical Rayleigh number at the onset 
of convection (\ref{Rac_Busse}).
The moderate variation of $\Pra$ in the database implies that it can
be omitted without significant loss in the quality of the fit
(see figure~\ref{Els_phys_fig_rolePr}.b). This provides an even simpler
scaling law, valid for the available range of $\Pra$
\begin{equation}
{\Lambda} \sim (\widetilde{R}-\widetilde{R}_d) \, \q \, ,
\qquad 
\mbox{with}\quad
\chirel=0.512\, .
\label{notrefit2}
\end{equation}
It involves input parameters only, and its derivation was guided by physical arguments.
Besides, it is worth noting that (\ref{notrefit2}) as well as (\ref{notrefit}) imply
a dependence of the magnetic field amplitude on the rotation rate
$\Omega$. This contradicts earlier claims of
saturation values independent on the rotation rate.

\begin{figure}
\centerline{\includegraphics[width=0.9\textwidth]{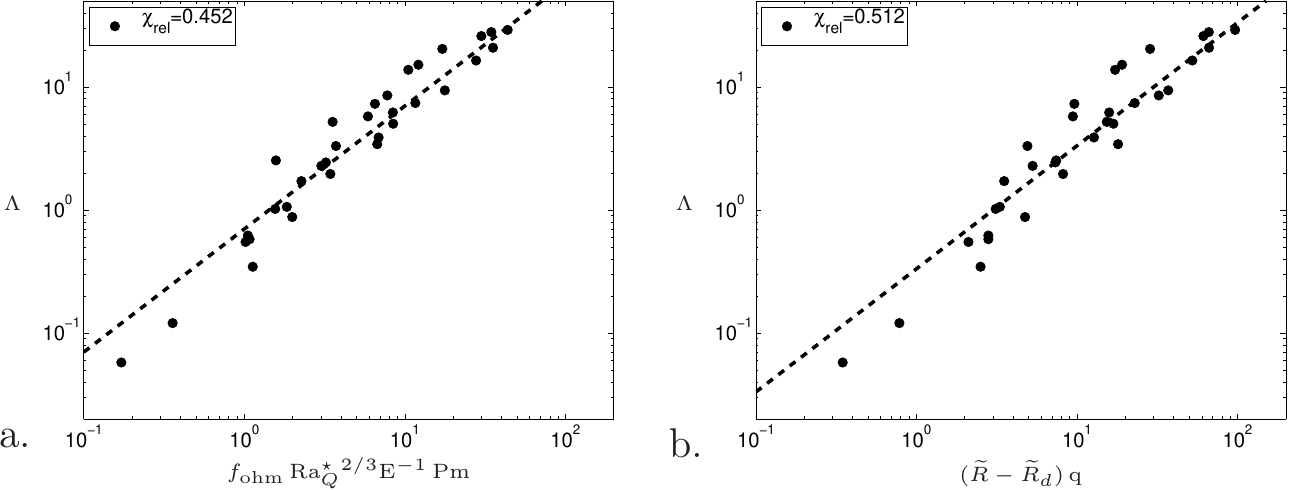}}
\caption{Comparison of the earlier power based scaling law \citep{Chr10} and 
our proposed predictive scaling law for the magnetic field strength: relations (a) (\ref{fitCA06_nonopt_mod}) 
and (b) (\ref{notrefit2}). Both graphs rely on a reduced $33$ dynamos database for which $\widetilde{R}_d$ and $\fohm$ are available.}
\label{final}
\end{figure}

Relations (\ref{fitCA06_nonopt_mod}) and (\ref{notrefit2}) are applied to
a reduced $33$ dynamos database (for which all quantities involved in both relations are available) 
and represented in figure~\ref{final}. The relative misfits given in
\eqref{fitCA06_nonopt_mod}, \eqref{notrefit}, \eqref{notrefit2} are
computed on the basis of this reduced database.

Note that the power based relation (\ref{fitCA06_nonopt_mod}) does not involve any distance to the onset of dynamo action. Indeed, 
the parameter ${\Ray_Q^\star}$ does not vanish at the onset of dynamo action (it vanishes at the onset of convection, 
see equation (\ref{def_RastarQ})). 
The parameter $\fohm$ however corrects this issue, as it tends to zero at the onset of dynamo action.

\section{Conclusion}

In this study, we combine a numerical approach, which consists in establishing scaling laws for quantities of 
interest thanks to a multiple linear regression method applied to numerical data under the approximation of power laws, 
and a physical approach based either on energetics or on forces balances. Our numerical approach is based on a $102$ dynamos
database (U. Christensen) corresponding to Boussinesq fully convecting ($\Nu>2$) and dipolar dynamo models.\\

In a first phase, we focus our attention on scaling laws for the magnetic field strength as a function of 
the injected power by buoyancy forces, quantified by the flux-based Rayleigh number $\Ray_Q^\star$. We show that 
the scaling laws previously obtained in the literature mainly correspond 
to the simple writing of the energy balance between production and 
dissipation, which is necessarily valid for any dynamo in statistical equilibrium. 
Such power based scaling laws are thus very general and applicable to any dynamo in statistical equilibrium
irrelevantly of the dynamo mechanism \citep[e.g.][]{SPD12}.

The description of the magnetic dissipation length scale $\ell_B$
determines the quality of the approximation. Assuming a constant value for
$\ell_B$ already provides a very good description of the numerical database. 
Improved fits can be obtained based on finer assumptions for $\ell_B$.
However, none of the proposed scaling laws corresponds to a realistic physically based
relation to describe the numerical database (see section~\ref{inter_phys}).

{The second part of our study aims at establishing predictive scaling laws 
(i.e. as a function of input parameters only) for the magnetic field strength. 
Our reasoning is guided by physical arguments such as forces balances, and the numerical 
database is only used to validate the proposed relations. Indeed, we have shown that scaling laws obtained through a direct numerical fit 
can be biased by the numerical sample. It is in particular the case for the Ekman and magnetic Prandtl numbers, 
whose ranges are strongly correlated in the database.
The flux-based Rayleigh number $\Ray_Q^\star$, which is a measured quantity, is replaced either by 
the normalised distance of the Rayleigh number to the onset of convection (denoted as $\widetilde{R}-1$) 
or by the normalised distance to the onset of dynamo action (measured by $\widetilde{R}-\widetilde{R}_d$). This last quantity is unfortunately 
only available for a subset of the numerical database. }

Our four control parameters are the Ekman number, the Prandtl number, the magnetic Prandtl number and the relative distance 
to the onset of convection (resp. dynamo action). Our reasoning follows four steps. 

The first step of the reasoning provides a scaling law for the magnetic field strength as a function of the distance to the onset of dynamo
in term of flow amplitude, which is $\Lambda \sim {\left(\Rm-\Rm_d\right)} \, {\Ek}^{1/3}$ and which matches numerical data.  
It is deduced from the balance between the Lorentz force and the viscous force associated to the 
flow distorsion \citep{Dormy07}. 

The second one consists in establishing the link between the injected power (measured by $\Ray_Q^\star$) and
$\widetilde{R}-1$, by using the definition of $\Ray_Q^\star$, the relation between $\Nu$ and $\widetilde{R}$ \citep[e.g.][]{King10} and 
previously established dependences of the critical Rayleigh number at the onset of convection on the 
Ekman and Prandtl numbers \citep{Busse70}. 

The third step deals with the derivation of a scaling law for the flow amplitude. 
The Viscous-Archimedean-Coriolis scaling \citep{King13} matches
the numerical data. Especially, the characteristic velocity length scale of the flow 
depends on $\Ek^{1/3}$ in numerical simulations, which proves that viscous effects play a non-negligible role in the bulk of the flow. 
The role of inertia is shown to be negligible on this length scale for dipolar dynamos compared to that of viscous effects.

Finally, in a fourth step, the combination of the aforesaid
  results leads to a surprisingly simple
predictive scaling law, that is 
$\Lambda \sim (\widetilde{R}-\widetilde{R}_d) \, \q \, (\Pra/(1+\Pra))^{2/3}$, which involves input parameters only, 
contrary to previous published scaling laws, and which properly describes
available numerical data (as stressed in the text, the $\Pra$ dependence is not tested by the database and can be omitted 
here without loss). This scaling law 
relies on the dominant forces balance in the numerical dynamos. 
Contrary to power based scaling laws, it is applicable in the parameter range covered 
by this study, but will not be satified in general (e.g. if inertial forces
become significant). Besides its predictive power, it also provides information on the underlying forces balance 
at work in the dynamo simulations.

Introducing predictive scaling
laws, based on control parameters only, allows to underline two important ideas. First, the present numerical
models do not operate in a dominant forces balance relevant for the
geodynamo. Indeed, viscous effects are shown to be essential and
extrapolation to geophysically relevant parameters produces strongly
underestimated amplitudes for the magnetic field. Secondly, it allows to demonstrate the 
clear dependence of the magnetic field strength on the rotation rate
$\Omega$. 

\newpage

\appendix

\section{The multiple linear regression approach}
\label{mlr}

As in previous studies \citep{CT04,Chr06,Stelzer13},
we restrict our scaling analysis to power laws of the form 
\begin{equation}
\mathbf{Y} \propto \alpha \prod_{j=1}^{p} {\mathbf{X}_j}^{\beta_j} \, ,
\label{regress}
\end{equation}
where $\mathbf{Y}$ is the n-dimensional vector of output data which we want
to fit, and $\mathbf{X_j}$ are the p n-dimensional predictor variables. 
Taking the logarithm of (\ref{regress}) transforms the model in a multiple linear regression problem
\begin{equation}
\mathbf{log(Y)}=\beta_0 + \beta_1\, \mathbf{log(X_1)} + \beta_2\,
\mathbf{log(X_2)} + ... + \beta_p\, \mathbf{log(X_p)} + \text{\boldmath $\varepsilon$} \,
. 
\label{Reglin}
\end{equation}
in which 
$\beta_0 = {\rm log} ( \alpha)$,
and {\boldmath $\varepsilon$} in an n-dimensional vector measuring the
misfit.

In the following, $\mathbf{log(Y)}$ is replaced by $\widetilde{\mathbf{Y}}$
and $\mathbf{log(X_j)}$ by $\mathbf{\widetilde{X}_j}$ for clarity. The system of n
equations (\ref{Reglin}) can be represented in matrix notation as  
\begin{equation}
\widetilde{\mathbf{Y}}=\widetilde{\mathbf{X}} \,\text{\boldmath $\beta$} + \text{\boldmath $\varepsilon$} \, ,
\end{equation}
where $\widetilde{\mathbf{X}}$ is refered to as the design matrix
$\left[\mathbf{I}~\mathbf{\widetilde{X}_1} ~...~ \mathbf{\widetilde{X}_p} \right]$ and {\boldmath $\beta$} is a $(p+1)$-dimensional 
vector containing the whole set of regression coefficients. The vector
{\boldmath $\beta$} can be estimated using least square estimates. The misfit {\boldmath $\varepsilon$} is assumed 
to follow a Gaussian centered distribution with a variance $\sigma$ which is assumed to 
be a constant. The corresponding fitted model is  
\begin{equation}
\hat{\widetilde{\mathbf{Y}}}=\widetilde{\mathbf{X}}\, \hat{\text{\boldmath $\beta$}} \, ,
\end{equation}
where 
\begin{equation}
\hat{\text{\boldmath $\beta$}}=\left(\widetilde{\mathbf{X}}^t
\cdot \widetilde{\mathbf{X}}\right)^{-1}\cdot \widetilde{\mathbf{X}}^t
\cdot\widetilde{\mathbf{Y}} \, .
\end{equation}
The variance $\sigma$ can be estimated by the unbiased estimator $\hat{\sigma}$ defined as
\begin{equation}
\hat{\sigma}^2=\frac{1}{n-p-1} \| \hat{\text{\boldmath $\varepsilon$}} \|^2 \, ,
\qquad 
\mbox{where}\quad
\hat{\text{\boldmath $\varepsilon$}}=\mathbf{Y}-\hat{\widetilde{\mathbf{Y}}}\, .
\end{equation}
As a measure of misfit between data and fitted values, we use as in \cite{Chr06} the mean relative misfit to the original data 
$y_{i}$ ($i \in \left(1,n\right)$), defined as
\begin{equation}
\mathbf{\chirel}=\sqrt{\frac{1}{n} \sum_{\rm i=1}^{n} \left(\frac{y_i-\hat{y}_i}{y_i}\right)^2}\,. 
\end{equation}
The estimator $\hat{\text{\boldmath $\beta$}}$ is unbiased and 
its covariance matrix can be estimated by
\begin{equation}
\hat{\text{\boldmath $\sigma$}}_{\hat{\beta}}^2= \hat{\sigma}^2 \left(\widetilde{\mathbf{X}}^t \cdot \widetilde{\mathbf{X}}\right)^{-1} \, ,
\end{equation} 
which is a $(p+1)\times(p+1)$ matrix. An estimation of the variance 
$\hat{\sigma}_{\hat{\beta}_j}$ of the $\hat{\beta}_j$ exponent ($j \in \left(0,p\right)$) is
\begin{equation}
\hat{\sigma}_{\hat{\beta}_j} = \hat{\sigma} \sqrt{\left(\left(\widetilde{\mathbf{X}}^t \cdot 
\widetilde{\mathbf{X}}\right)^{-1}\right)_{jj}} \, ,
\end{equation}
and the estimator $(\hat{\beta}_j-{\beta}_j)/\hat{\sigma}_{\hat{\beta}_j}$ follows a Student distribution \citep{Student08,Fisher25} with $(n-p-1)$ degrees of freedom. For the analysis performed in this article, $(n-p-1) \approx 100$. In
that case, the coefficient $\beta_j$ is comprised in the $95\%$ confidence interval
\begin{equation}
{\beta}_j = {\hat{\beta}_j} \pm 2\,\hat{\sigma}_{\hat{\beta}_j}.
\end{equation}
This method provides the following power law for $y$
\begin{equation}
y = \exp(\hat{{\beta}_0} \pm 2\,\hat{\sigma}_{\hat{\beta}_0}) \prod_{j=1}^{p} {x_j}^{\hat{\beta_j} \, \pm \, 2\,\hat{\sigma}_{\hat{\beta}_j} } \, ,
\end{equation}
which can be rewritten as 
\begin{equation}
y = \left(\exp(\hat{{\beta}_0})\cosh(2\,\hat{\sigma}_{\hat{\beta}_0}) \pm  \exp(\hat{{\beta}_0})\sinh(2\,\hat{\sigma}_{\hat{\beta}_0})\right) \prod_{j=1}^{p} {x_j}^{\hat{\beta_j} \, \pm \, 2\,\hat{\sigma}_{\hat{\beta}_j} } \, .
\end{equation}
In the present paper, the confidence intervals are provided in separated tables.

In a geometric interpretation where the essential quantity is reported in ordinate
and the optimal combination of fitting parameters in abscissa, the mean relative misfit $\chirel$ measures 
the relative ordinate distance between observations and estimations, without taking the abscissa 
distance into account. That is why its use is restricted to comparisons of fits for the same quantity $y$.
Besides, the mean relative misfit $\chirel$ is obviously expected to decrease 
with the number $p$ of predictor variables. As the system of equations 
(\ref{eq_NS}-\ref{eq_div}) is governed by four non-dimensional 
parameters ($\Ray$, $\Ek$, $\Pm$ and $\Pra$), the maximum number $p_{\rm max}$ of independent predictor variables is equal to $4$.
For further discussions on fitting errors, we refere the reader to \cite{Stelzer13}.

\section{Role of the fraction of ohmic dissipation $\fohm$ in empirical scaling laws for the magnetic field strength}
\label{Appfohm}

We stress here the pifalls of direct numerical fits, free from physical
insight, by showing that different a priori hypothesis on $\fohm$
yield contradictory results.

\subsection{Power based scaling laws derived from multiple linear regressions}
\label{section_fohm}

As noted in section~\ref{ProdDiss},
$\fohm$ is not at all a trivial 
parameter, as it involves both controlled and measured quantities.
Indeed with our notations, equation (\ref{def_f_ohm}) can be rewritten as 
\begin{equation}
\fohm = \left(1+ \frac{\Ro^2}{\Lo^2} \,\frac{{\ell_B^\star}^2}{{\ell_u^\star}^2}\,
\Pm\right)^{-1} \, ,
\label{eqfohm}
\end{equation}
where we introduced a kinematic dissipation length scale $\ell_u$ ($\ell_u^\star=\ell_u/L$), defined using time averaged quantities as
\begin{equation} 
\ell_u ^2\equiv \frac{\int_V \mathbf{u}^2 \,\dd V}{\int_V (\mathbf{\nabla} \times \mathbf{u})^2\, \dd V}
=2 \, \nu
 \frac{E_{kin}}{ D_{\nu}} \, .
\label{ldissu}
\end{equation} 
The main distinction between the scaling laws (\ref{fitCA06_nonopt}) and
(\ref{eq9Dav_adim}) respectively proposed by \cite{Chr06} and \cite{Davidson13} relies 
on the different exponent of $\fohm$. 

\begin{figure}
\centerline{\includegraphics[width=1\textwidth]{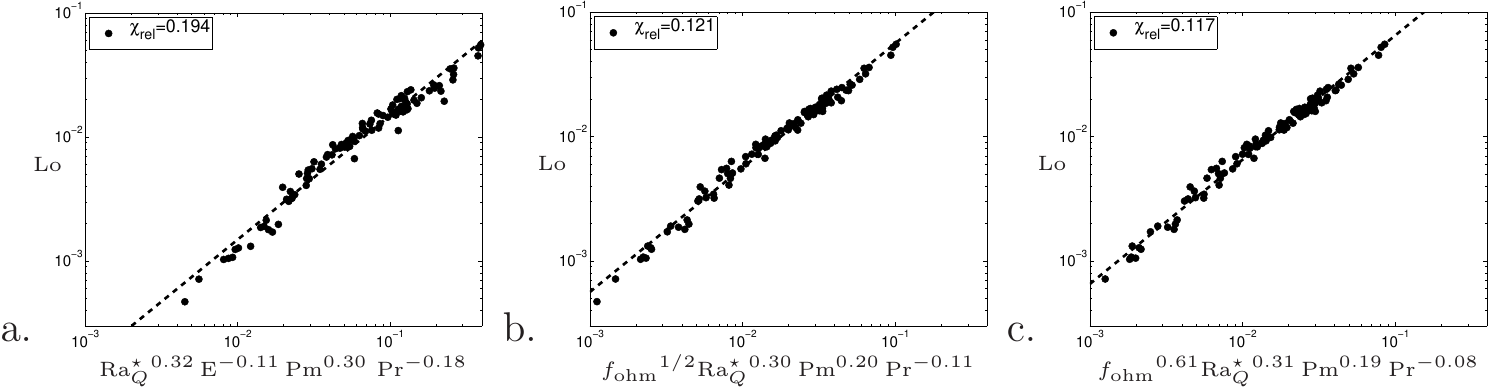}}   
\caption{The Lorentz number versus a combination of flux-based Rayleigh number, Ekman number, Prandtl number, 
magnetic Prandtl number (a) with no $\fohm$ dependence (equation (\ref{eqLo_sansfohm})), (b) with an additional 
fixed $\fohm ^{1/2}$ factor (equation (\ref{bestfitLo_3para})) and (c) an additional $\fohm$ dependence with an 
optimised exponent (equation (\ref{notrebestfitLo_5para})). This figure relies on the $102$ dynamos database.} 
\label{Lo_bestfit}
\end{figure}

First considering the best empirical scaling law for the magnetic field
strength in our database,
ignoring the $\fohm$ parameter, we get
\begin{equation}
\Lo \simeq 0.16 \, {\Ray_Q^\star}^{0.32} \, \Ek^{-0.11} \, \Pm^{0.30} \, \Pra^{-0.18}\, ,
\qquad 
\mbox{with}\quad
\chirel=0.194\, ,
\label{eqLo_sansfohm}
\end{equation}
(see figure~\ref{Lo_bestfit}.a). 
Note that in the above expression, the right-hand-side vanishes at the
onset of convection and not at the onset of dynamo action. This expression
is therefore obviously not valid close to the onset of dynamo action.

The balance between energy production and dissipation provides an exponent $1/2$ for $\fohm$ (see section~\ref{ProdDiss}).
The best power law approximation for $\Lo$ as a function of
$\Ray_Q^\star$ obtained by setting the exponent of $\fohm$ to $1/2$ is then
\begin{equation}
\Lo \simeq 0.78 \,\fohm^{1/2} \, {\Ray_Q^\star}^{0.32} \, , 
\qquad 
\mbox{with}\quad
\chirel=0.256 \, ,
\label{bestfitLo_f1_1para}
\end{equation}
whereas allowing for a dependence on $\Pm$ leads to
\begin{equation}
\Lo \simeq  0.64 \,\fohm^{1/2} \, {\Ray_Q^\star}^{0.31}\, \Pm^{0.17}\, ,
\qquad 
\mbox{with}\quad
\chirel=0.141 \, .
\label{bestfitLo_2para}
\end{equation}
These two expressions correspond to the fits (\ref{fitCA06_nonopt}) and (\ref{fitCA06}) of \cite{Chr06}.
The exponents do not exactly match because the numerical database used here is somewhat larger. 
However, the two latter relations are rigorously recovered if we apply 
our algorithm to the $65$ dynamos numerical database of \cite{Chr06}. This validates
the multiple linear regression approach used in the present paper.
The role of the parameter $\Pra$ is found to be negligible using the $65$ dynamos numerical 
database of \cite{Chr06}. But the $102$ dynamos database used here contains more simulations 
corresponding to $\Pra \ne 1$ than the earlier \cite{Chr06} database ($32$ versus $17$). 
Considering an additional dependence on $\Pra$ yields
\begin{equation}
\Lo \simeq 0.56 \,  \fohm^{1/2} \,  {\Ray_Q^\star}^{0.30} \,\Pm^{0.20}\, \Pra^{-0.11}\, ,
\qquad 
\mbox{with}\quad
\chirel=0.121 \, ,
\label{bestfitLo_3para}
\end{equation}
(see figure~\ref{Lo_bestfit}.b), where the dependence on $\Pra$ is not negligible. On the contrary, 
the contribution of the Ekman number appears negligible (taking $\Ek$ into account only provides a 
very minor improvement of $\chirel$ and a small power of $\Ek$).

It is however natural in a fitting approach to let the exponent $\fohm$ be
determined by the multiple linear regression approach.
Moreover, as noted above, the $\fohm$ parameter is usually argued to be
equal to unity in planetary dynamos.
The best power law with the above parameters is
\begin{equation}
\Lo \simeq 0.66 \,{\fohm^{0.61} \,\Ray_Q^\star}^{0.31}\, \Pm^{0.19}\, \Pra^{-0.08} \, ,
\qquad 
\mbox{with}\quad
\chirel=0.117\, ,
\label{notrebestfitLo_5para}
\end{equation}
(see figure~\ref{Lo_bestfit}.c). The contribution of $\Ek$ is negligible, this last relation thus involves 
five non-dimensional parameters only, which corresponds to the maximum number of 
independent parameters in the problem (see appendix~\ref{mlr}). 
Table~\ref{incertitudes_adim1} gathers the fitted values corresponding to equations (\ref{eqLo_sansfohm}), (\ref{bestfitLo_f1_1para}), 
(\ref{bestfitLo_2para}), (\ref{bestfitLo_3para}) and (\ref{notrebestfitLo_5para}) including their $95\%$ confidence interval. 
The exponents in relation (\ref{notrebestfitLo_5para}) 
are not significantly different from those in relation (\ref{bestfitLo_3para}). In particular, the $95\%$ confidence interval associated to the 
optimised value $0.61$ of the exponent of $\fohm$ in (\ref{notrebestfitLo_5para}) includes the value $1/2$ provided 
by the energy balance.

The relative error on the exponents of $\Ray_Q^{\star}$, $\Pm$ and $\fohm$
is in general moderate (less than $15\%$). 
The error for the estimation of the exponent of $\Pra$ is more important (between $30\%$ and $50\%$): the distribution 
of the control parameter $\Pra$ in our dataset, although wider than in the dataset used in \cite{Chr06}, 
is indeed not wide enough to establish a clear dependence on $\Pra$. The parameter $\Ek$ 
appears only in relation (\ref{eqLo_sansfohm}) where the output parameter $\fohm$ is not taken into account, 
with a relative error of $50\%$ for the corresponding exponent. 

Finally, note that equations (\ref{eqLo_sansfohm}), (\ref{bestfitLo_3para}) and (\ref{notrebestfitLo_5para}) can be related by 
introducing the best power law approximation for $\fohm$ as a function of
$\Ray_Q^\star$, $\Ek$, $\Pm$ and $\Pra$, i.e.
\begin{equation}
\fohm \simeq 0.07 \, \Ek^{-0.17} \, \Pm^{0.18} \, \Pra^{-0.18} \, ,
\qquad 
\mbox{with}\quad
\chirel=0.249\, ,
\label{fohm}
\end{equation}
where the contribution of $\Ray_Q^\star$ is found to be negligible ($95\%$ confidence intervals in table~\ref{incertitudes_adim1}). 
The high corresponding relative misfit ($25\%$) reveals that 
the dependence of $\fohm$ on other parameters can not be reliably approximated by a simple power law.

\begin{table}
\footnotesize
\centering
\begin{tabular}{c c c c c c c c} 
\hline
 &  Pre-factor & $\fohm$ & $\Ray_Q^{\star}$ & $\Ek$ & $\Pm$ & $\Pra$ & $\chirel$  \\
\hline
$\Lo$ &  $0.157 \pm 0.050$  & $\times$ & $0.318 \pm 0.027$ & $-0.111 \pm 0.053$ & $0.295 \pm 0.039$ & $-0.176 \pm 0.056$ & $0.194$ \\
$\Lo$ & $0.777 \pm 0.168$ & $1/2$ & $0.322 \pm 0.017$ & $\times$ & $\times$ & $\times$ & $0.256$ \\
$\Lo$ & $0.638 \pm 0.080$ & $1/2$ & $0.313 \pm 0.009$ & $\times$ & $0.167 \pm 0.023$ & $\times$ & $0.141$ \\
$\Lo$ & $0.561 \pm 0.063$ & $1/2$ & $0.302 \pm 0.009$ & - & $0.197 \pm 0.021$ & $-0.106 \pm 0.033$ & $0.121$ \\
$\Lo$ &  $0.661 \pm 0.114$ & $0.605 \pm 0.091$ & $0.309 \pm 0.010$ & - & $0.186 \pm 0.023$ & $-0.080 \pm 0.039$ & $0.117$ \\
$\fohm$ & $0.073 \pm 0.026$ & $\times$ & - & $-0.170 \pm 0.030$ & $0.180 \pm 0.042$ & $-0.178 \pm 0.054$ & $0.249$ \\
\hline
\end{tabular}
\caption{Optimal scaling laws obtained by the multiple linear regression method, for $\Lo$ and $\fohm$ ($95\%$ confidence intervals): 
relations (\ref{eqLo_sansfohm}), (\ref{bestfitLo_f1_1para}), (\ref{bestfitLo_2para}), (\ref{bestfitLo_3para}), 
(\ref{notrebestfitLo_5para}) and (\ref{fohm}).}
\label{incertitudes_adim1}
\normalsize
\end{table}

\subsection{Extrapolation to natural dynamos}
\label{natural}

The results of appendix~\ref{section_fohm} deserve careful analysis.
Equation (\ref{notrebestfitLo_5para}) may be indeed viewed as a minor 
improvement in the quality of the fit, resulting from the introduction of
an additional degree of freedom in the problem.
Besides, the $\fohm$ parameter involves most of the quantities we are
trying to fit, that is to say $\Lo$, $\Ro$, $\ell^\star_B$ and $\ell^\star_u$ (see
equation (\ref{eqfohm})). 

Nevertheless, the above study clearly indicates that different
scaling laws can be proposed for $\Lo$, depending on exponents considered for $\fohm$. 
If one adopts the usual assumption that  $\fohm=1$ for planetary
applications, the resulting relation for such applications will not depend on the exponent of $\fohm$.
To illustrate this, we can write equations
(\ref{eqLo_sansfohm}), (\ref{bestfitLo_3para}) and
(\ref{notrebestfitLo_5para}) in dimensional form assuming $\fohm=1$. These are respectively 
\begin{equation}
\B \sim \mu^{1/2}\,\, \Power^{0.32}\,\, \Omega^{0.16}\,\, L^{-0.37}\, \rho^{0.18} \,\nu^{0.01} \,\eta^{-0.30}\, \kappa^{0.18}  \, ,
\qquad 
\mbox{with}\quad
\chirel= 0.194\, ,
\label{Bdim1}
\end{equation}
\begin{equation}
\B \sim \mu^{1/2}\,\, \Power^{0.30}\, \,\Omega^{0.09}\,\, L^{-0.51}\,\, \rho^{0.20}\,\,
\nu^{0.09} \,\,\eta^{-0.20}\,\, \kappa^{0.11} \, ,
\qquad 
\mbox{with}\quad
\chirel=0.121\, ,
\label{Bdim2}
\end{equation}
and 
\begin{equation}
\B \sim \mu^{1/2}\,\, \Power^{0.31} \,\,\Omega^{0.07} \,\,L^{-0.55}\,\,
\rho^{0.19}\,\, \nu^{0.11}\,\, \eta^{-0.19} \, \,\kappa^{0.08} \, ,
\qquad 
\mbox{with}\quad
\chirel=0.117\, .
\label{Bdim3}
\end{equation}
Table~\ref{incertitudes_dim1} gathers the above fitted values and the corresponding confidence intervals. The latter are calculated 
using the $95\%$ confidence intervals found in the non-dimensional scaling laws, considering their more pessimistic 
combination. By this process, the three relations can not be distinguished: for the exponents of all parameters, there exists an 
interval common to the three expressions. But if we consider $70\%$ confidence intervals as Stelzer \& Jackson did, the incertitude of 
the exponent $\hat{\beta}_j$ is equal to $\hat{\sigma}_{\hat{\beta}_j}$ instead of $2\,\hat{\sigma}_{\hat{\beta}_j}$ 
(see appendix~\ref{mlr}). We can also deduce that expression (\ref{Bdim1}) predicts a dependence of 
$\B$ on $\Omega$ which is twice that of (\ref{Bdim3}). A similar effect can be noted for the dependence on $\kappa$. The dependence 
on $\eta$ predicted by (\ref{Bdim1}) is also $1.5$ higher than that predicted by the scaling law (\ref{Bdim3}). Finally, 
(\ref{Bdim1}) predicts a much weaker dependence on $\nu$ than (\ref{Bdim3}) ($1/10^{\rm th}$ factor).
Thus, in the limit of $70\%$ confidence intervals, the dependence of the magnetic field strength on physical parameters seems
to depend on the role given to $\fohm$ in the numerical fit. 

\begin{table}
\scriptsize
\centering
\begin{tabular}{c c c c c c c c c} 
\hline
  & $\Power$ & $\Omega$ & $L$ & $\rho$ & $\nu$ & $\eta$ & $\kappa$ & $\chirel$  \\
\hline
$\B/\mu^{1/2}$  & $0.318 \pm 0.027$ & $0.157 \pm 0.134$ & $-0.368 \pm 0.241$ & $0.182 \pm 0.027$ & $0.008 \pm 0.142$ & $-0.295 \pm 0.039$ & $0.176 \pm 0.056$ & $0.194$ \\
$\B/\mu^{1/2}$  & $0.302 \pm 0.009$ & $0.094 \pm 0.027$ & $-0.510 \pm 0.045$ & $0.198 \pm 0.009$ & $0.091 \pm 0.054$ & $-0.197 \pm 0.021$ & $0.106 \pm 0.033$ & $0.121$ \\
$\B/\mu^{1/2}$  & $0.309 \pm0.010$ & $0.073 \pm 0.030$ & $-0.545 \pm 0.050$ & $0.191 \pm 0.010$ & $0.106 \pm 0.062$ & $-0.186 \pm 0.023$ & $0.080 \pm 0.039$ & $0.117$ \\
\hline
\end{tabular}
\caption{$95\%$ confidence intervals associated to exponents in the dimensional scaling laws for the 
magnetic field strength corresponding to relations (\ref{Bdim1}), (\ref{Bdim2}) and (\ref{Bdim3}).}
\label{incertitudes_dim1}
\normalsize
\end{table}

Using an estimate for the Earth's core of $\Ray_{Q \, \,\rm Earth}^{\star}=10^{-14}$ \citep[e.g.][]{Chr06} in 
(\ref{eqLo_sansfohm}), (\ref{bestfitLo_3para}) and (\ref{notrebestfitLo_5para}) yields $\B_{\,\rm Earth}=0.10$ mT, 
$\B_{\,\rm Earth}=0.05$ mT and $\B_{\,\rm Earth}=0.05$ mT respectively.
It should be compared to the rms magnetic field strength inside the Earth's core, 
estimated to be of the order of $2-4$~mT
\citep[e.g. see][]{Buffett10,Gillet10}. 
Our values above are lower than this estimated value by a factor $20$-$40$,
just as the values obtained by \cite{Chr06} and \cite{Stelzer13}.

\section{The magnetic dissipation length scale $\ell_B$}
\label{Applengthlb}

\subsection{The $\ell_B$ length scale as a function of the flow amplitude}
\label{section_lB_Rm}

We interpret here earlier scaling laws in terms of assumptions made on $\ell_B^\star$ and their implications for $\Ro$ and $\ell^\star$. 

\begin{figure}
\centerline{\includegraphics[width=0.9 \textwidth]{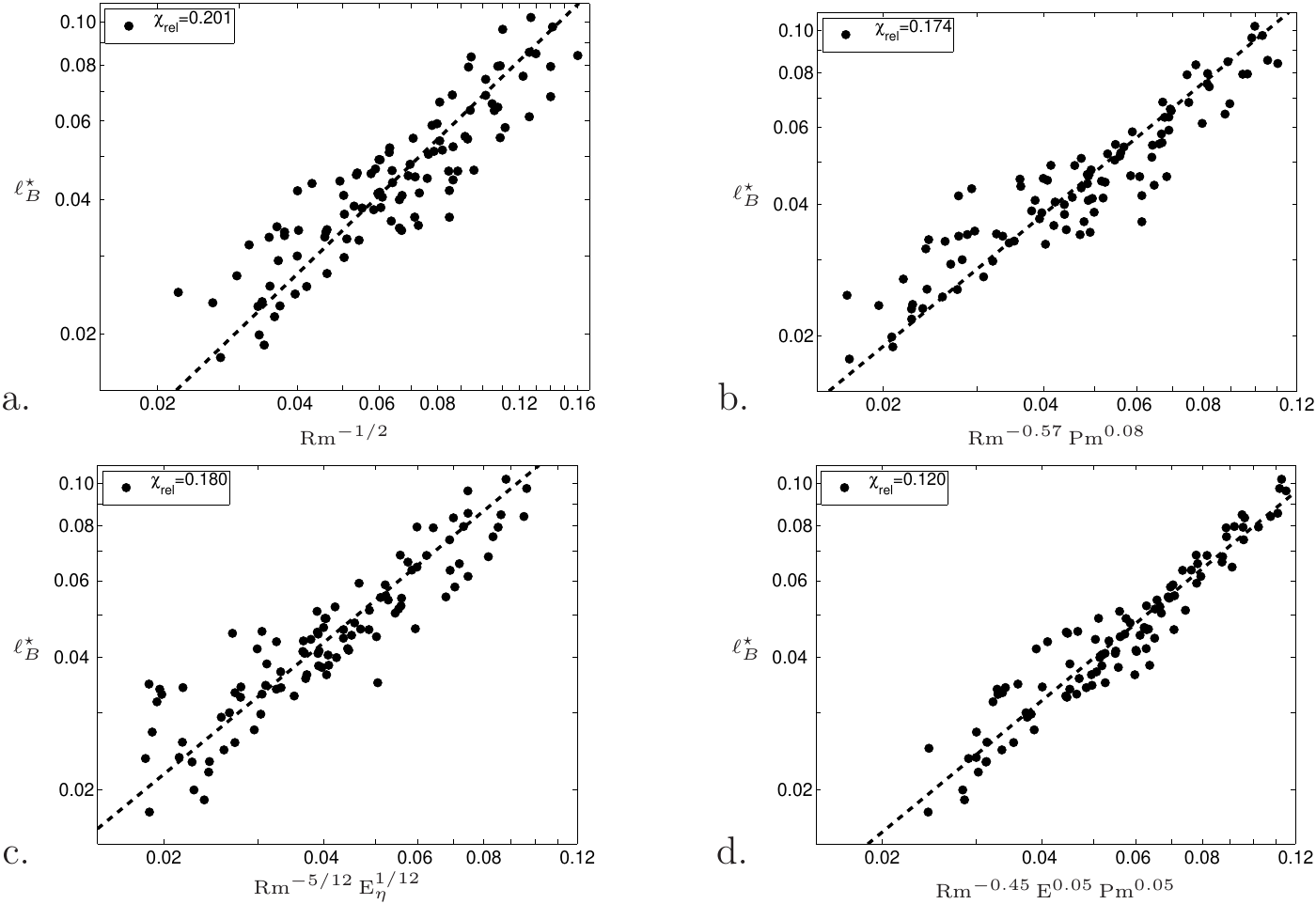}}
\caption{The magnetic dissipation length scale versus a combination of the magnetic Reynolds number, 
the Ekman number and the magnetic Prandtl number: (a) ${\ell_B^\star} \sim \Rm^{-1/2}$ (b) ${\ell_B^\star} \sim
\Rm^{-0.57}\, \Pm^{0.08}$ \citep[both derived from][]{CT04} (c) ${\ell_B^\star} \sim \Rm^{-5/12}
 \, \Ek_{\eta}^{1/12}$ \citep{Chr10} (d) ${\ell_B^\star} \sim \Rm^{-0.45} \,\Ek^{0.05}\, \Pm^{0.05}$ \citep{Stelzer13}. 
These graphs rely on the full $102$ dynamos database.} 
\label{FigfitslB}
\end{figure}

\cite{CT04} have 
empirically shown that $\tau_{\eta}^\star \sim \Rm^{-1}$~. Because  
\(\tau_{\eta}^\star \sim {\ell_B^{\star}}^2\)~, this provides
${\ell_B^\star} \sim \Rm^{-1/2}$ \citep[see figure~\ref{FigfitslB}.a, see also][]{RobertsKing13}. According to equation (\ref{eql}), 
this scaling law corresponds to assuming that $\ell^\star \sim 1$, i.e. $\ell$ is the width of the spherical shell. 
It is reasonably consistent with the $102$ dynamos database used in this paper, since 
$\Rm~{\ell_B^\star}^2$ varies from \(0.19\) to \(1.25\), that is to say over about one order of
magnitude. Moreover, note that some of the values are higher than unity: it is symptomatic of the role played by correlations
between the norm and direction of $\bfu$ and
$\bfB$.
\cite{CT04} have empirically
improved the above scaling law to $  {\ell_B^{\star}} \sim 
\Rm^{-0.49}\, \Rey^{-0.08}$, where $\Rey$ is the Reynolds number
($\Rey=\Rm\, \Pm^{-1}$). This expression can be reformulated as ${\ell_B^\star} \sim
\Rm^{-0.57}\, \Pm^{0.08}$ (see figure~\ref{FigfitslB}.b). 
Thanks to a larger numerical data sample, this last scaling law
has been optimised by \cite{Chr10} as ${\ell_B^\star} \sim \Rm^{-5/12}
 \, \Ek_{\eta}^{1/12}$ (see figure~\ref{FigfitslB}.c), and then by \cite{Stelzer13} as ${\ell_B^\star} \sim \Rm^{-0.45} \,\Ek^{0.05}\, \Pm^{0.05}$ 
(see figure~\ref{FigfitslB}.d). 

As expected, the relative misfit $\chirel$ decreases when the
number of predictor variables increases. Moreover, note that fits in
figure~\ref{FigfitslB} are based on $102$ numerical simulations extracted 
from the data sample provided by U. Christensen. Thus, the sample used in figure~\ref{FigfitslB} is 
larger than the one originally used by \cite{CT04} and \cite{Chr10}, and slightly different from that
used by \cite{Stelzer13} (also based on the $185$ dynamos database of 
U. Christensen but including $\fdip>0.35$ dynamos).

Finally, whereas the simple scaling law used in figure~\ref{FigfitslB}.a corresponds
to a simple physical assumption on the length scale $\ell$, the three other laws,
albeit more accurate, are simply based on empirical fits.

\subsection{The $\ell_B$ length scale as a function of the injected power}
\label{range_ra_star_Q}
Whereas the four aforesaid scaling laws rely on the magnetic Reynolds
number, scaling laws for the magnetic field amplitude based on a
production/dissipation balance rely on the flux-based Rayleigh number
\(\Ray_Q^\star\) (see sections \ref{ProdDiss}-\ref{Power_section}). It is 
therefore natural to seek for relations between the dissipation length scale \( \ell_B^\star\) 
and \(\Ray_Q^\star\)~.

Indeed, published scaling laws for the amplitude of the magnetic field,
such as the empirical scaling laws of \cite{Chr06} (see our equations
\eqref{fitCA06_nonopt}
and \eqref{fitCA06}),
can readily be translated in terms of scaling laws for \(\ell_B^\star\)~.
Thus, using equations (\ref{eqLoadim}) and (\ref{Power_CA06}), the
scaling laws (\ref{fitCA06_nonopt}) and (\ref{fitCA06}) respectively imply  
\(
{\ell_B^\star} \sim {\Ray_Q^\star}^{-0.16}\, \Ek_{\eta}^{1/2} \, ,
\)
and
\(
{\ell_B^\star} \sim {\Ray_Q^\star}^{-0.18}\, \Ek^{1/2} \, \Pm^{-0.39}\, .
\) 
It is interesting to note that in the representations of relations 
(\ref{fitCA06_nonopt}) and (\ref{fitCA06}) by \cite{Chr06}, the x-coordinate varies over six orders 
of magnitude \citep[see figures~\ref{FigloiUCR}.a,b in the present paper 
and figures~8-9 in][]{Chr06} while none of the control parameters varies over such a wide range. 
Thus, figure~\ref{FigCA06} (i.e. the above two relations) offers a somewhat
more challenging representation of the very same expressions
(\ref{fitCA06_nonopt}) and (\ref{fitCA06}) in so far as the axes vary on a smaller range.

The above scaling laws expressing $\ell_B^\star$ as a function of \({\Ray_Q^\star}\) were
deduced from (\ref{fitCA06_nonopt}) and (\ref{fitCA06}). As
\(\ell_B^\star\) is related to both $\ell^\star$ and $\Ro$, they also imply
relations between these two parameters and \({\Ray_Q^\star}\).
It is through these relations that the first of these scaling laws for the
magnetic field strength was originally physically interpreted by
\cite{Chr06}, \cite{Chr10} and \cite{Jones11}
(see section~\ref{inter_phys}).

\section{Length scales in Davidson's (2013) demonstration}
\label{AppDav}
The magnetic dissipation length scale denoted as $\ell_B$ in the present
paper is refered to as $\ell_{\rm min}$ in \cite{Davidson13}. 
Besides, he carefully introduced two length scales $\ell_{\parallel}$ and $\ell_{\perp}$ (the integral
length scales parallel and perpendicular to the rotation axis). The length scale $\ell_{\parallel}$ can be approximated by $L$, 
and the length scale $\ell_{\perp}$ corresponds to $\ell_u$ introduced in the present paper in 
appendix~\ref{section_fohm}. \cite{Davidson13} is interested in planetary dynamos, for which $\fohm \simeq 1$. We consider here the
question of the applicability of his analytical results to the length scales computed from the numerical database. 

Davidson's dimensional analysis leading to relation (\ref{Davidson1}) is based on the assumption that $\ell_B^2/\eta$ 
is independent on the rotation rate. This assumption, which was made in the limit relevant to planetary interiors,
does not seem to extend to the parameter regime of numerical models. 
Indeed, using (\ref{eqB}), the scaling law (\ref{Bdim1}) for the magnetic field strength implies
\begin{equation}
\frac{\ell_B^2}{\eta} \sim  \Omega^{0.314 \pm 0.268} \, ,
\label{lBdim1}
\end{equation}
($95\%$ confidence interval). Admittedly, the relative confidence interval is large, but the 
non-dependence of $\ell_B^2/\eta$ on the rotation rate, while sensible in the regime relevant to the geodynamo,
 is not relevant to the numerical data used here. 

\begin{figure}
\centerline{\includegraphics*[width=0.9\textwidth]{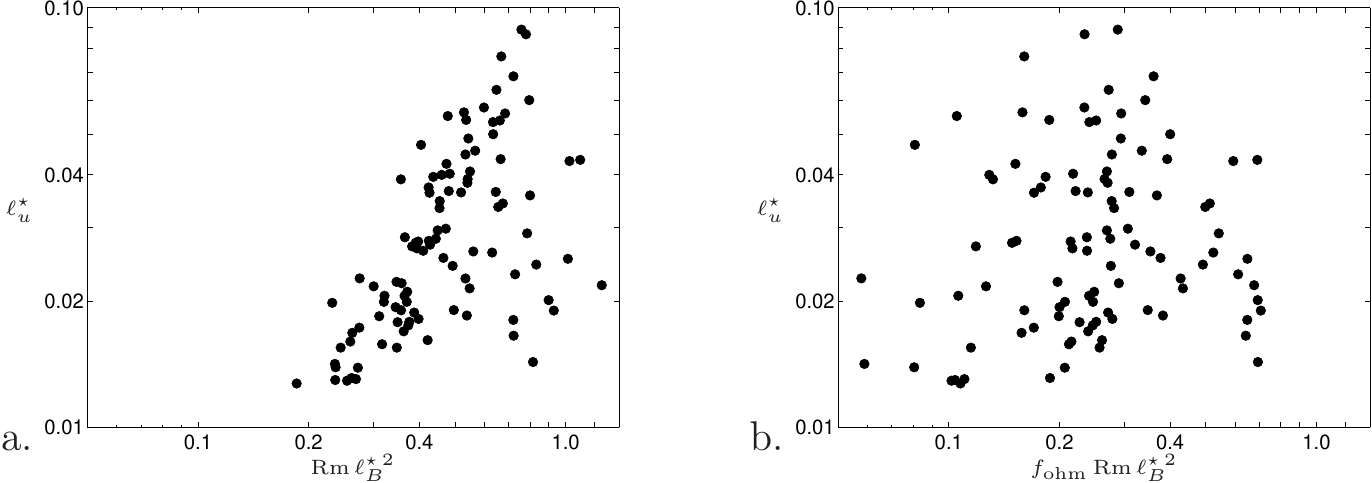}}
\caption{The length scale $\ell_u^\star$ as a function of (a) $\Rm\,{\ell_B^\star}^2$ (relation (\ref{Dav1})) 
(b) $\fohm\,\Rm\,{\ell_B^\star}^2$ (relation (\ref{Dav2})). These two graphs rely on the $102$ dynamos database. 
This figure highlights that the hypothesis made by \cite{Davidson13}, although well suited for planetary dynamos, 
are not met by numerical models.}
\label{lu_verifDav}
\end{figure}

Besides, neglecting viscous effects, he considered the balance of the curl of the
Coriolis force, the buoyancy force and the Lorentz force (the so-called MAC-balance, i.e. his equation~(10)). Its combination with
relation (\ref{Davidson1}) provided by his dimensional analysis leads to $\ell_B^2 \sim \eta u^{-1}
\ell_{\perp}$, which can be rewritten in its non-dimensional form as 
\begin{equation}
\ell_{\perp}^\star \sim  \Rm  \, {\ell_B^\star}^2  \, .
\label{Dav1}
\end{equation} 
By comparison with equation (\ref{eql}), that means 
that $\ell_{\perp}$ corresponds to our length scale $\ell$ defined in section~\ref{lengthb}. The distinction between $\ell$ and 
$\ell_{\perp}$ is proved important in our study. Figure~\ref{lu_verifDav}.a highlights that they should not be confused. 
The two assumptions, $\fohm \simeq 1$ and negligible viscous effects, are indeed not verified in numerical experiments.
 
If we use equation (\ref{eq9Dav}) \citep[equation~(9) in][]{Davidson13} rather than (\ref{Davidson1}) 
\citep[equation~(6) in][]{Davidson13} to take $\fohm$ 
into account, we get $\ell_B^2 \sim \eta \, \fohm^{-1}\, u^{-1} \, \ell_{\perp}$,
which can be rewritten in its non-dimensional form as 
\begin{equation}
\ell_{\perp}^\star \sim  \, \fohm \,  \Rm  \, {\ell_B^\star}^2  \, .
\label{Dav2}
\end{equation} 
This expression corresponds to a modified form of relation (\ref{Dav1}), adapted to $\fohm<1$ cases. 
Figure~\ref{lu_verifDav}.b shows that even such an $\fohm$ dependence does not provide a good description of 
the numerical data. This confirms that the assumption of 
negligible viscous effects, valid in the bulk of the Earth's core, is not applicable to present 
numerical simulations. Davidson's study therefore relies on 
assumptions relevant to the geodynamo, but not to present direct numerical simulations.

\section{Estimation of the onset of dynamo action}
\label{AppSeuil}
\begin{figure}
\centerline{\includegraphics*[width=0.9\textwidth]{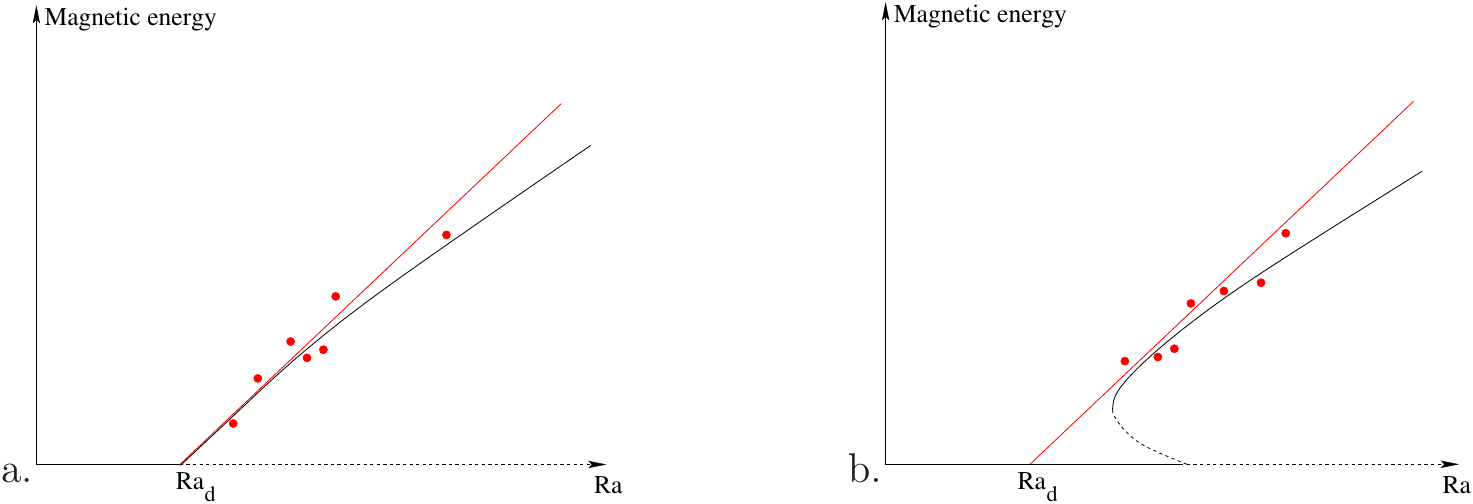}}
\caption{Schematic representation of the behaviour of the magnetic energy as a function of the Rayleigh number, for a (a) supercritical and (b) 
subcritical dynamo bifurcation. The solid (resp. dashed) lines indicate stable (resp. unstable) branches \citep[see][]{Morin09}.
The linear interpolation (red solid line) associated to data (red points) provides the value of $\Ray_d$ in both cases.}
\label{esti_seuil}
\end{figure}

The critical values at the onset of dynamo action $\Ray_d$ gathered in table~\ref{dynamo_seuil} 
have been estimated through a linear interpolation of the magnetic energy as a function of the Rayleigh number near 
the onset of dynamo action (see section~\ref{control}). As underlined by \cite{Morin09}, 
the dynamo bifurcation can be either supercritical or subcritical (or take the form of isola), 
the nature of the bifurcation depending on the parameters. The estimation of the critical Rayleigh number 
in the former case is represented in figure~\ref{esti_seuil}.a: the linear interpolation of data near the dynamo 
threshold provides $\Ray_d$. In the case of subcritical bifurcations, the critical $\Ray_d$ estimated by our 
method corresponds to the continuation of the subcritical branch, as shown in figure~\ref{esti_seuil}.b. A similar approach is 
used to determine $\Rm_d$.

\section*{Acknowledgments} 
The authors want to acknowledge stimulating discussion with Peter Davidson and gratefully thank Uli Christensen 
for sharing his numerical database.

\bibliographystyle{biblio_style}
\bibliography{biblio}

\end{document}